\documentclass{article}

\usepackage{color}

\def\a{\alpha}
\def\b{\beta}

\def\ga{\gamma}
\def\de{\delta}   

\def\phi{\varphi}
\def\la{\lambda}

\def\s{\sigma}

\def\vth{\vartheta}

\def\F{{\mathcal F}}
\def\G{{\mathcal G}}

\def\R{{\bf R}}

\def\X{{\mathcal X}}
\def\Y{{\mathcal Y}}
\def\T{{\mathrm T}}

\def\Ga{\Gamma}
\def\De{\Delta}
\def\La{\Lambda}

\def\pa{\partial}

\def\d{{\rm d}}       

\def\ss{\subset}

\def\EOR{\hfill {$\odot$}}
\def\EOP{\hfill {$\triangle$}}

\def\({\left(}
\def\){\right)}
\def\[{\left[}
\def\]{\right]}
\def\=#1{\overline #1}

\def\~#1{\widetilde #1}
\def\wt#1{\widetilde #1}
\def\.#1{\dot #1}
\def\^#1{\widehat #1}

\def\mapright#1{\smash{\mathop{\longrightarrow}\limits^{#1}}}
\def\mapdown#1{\Big\downarrow\rlap{$\vcenter{\hbox{$\scriptstyle#1$}}$}}
\def\mapleft#1{\smash{\mathop{\longleftarrow}\limits^{#1}}}

\def\mapleft#1{\smash{\mathop{\longleftarrow}\limits^{#1}}}

\def\interno{\hskip 2pt \vbox{\hbox{\vbox to .18
truecm{\vfill\hbox to .25 truecm
{\hfill\hfill}\vfill}\vrule}\hrule}\hskip 2 pt}

\def\beq{\begin{equation}}
\def\eeq{\end{equation}}
\def\eqref#1{(\ref{#1})}

\def\symmref{EMS1,CGbook,KrV,Olv1,Olv2,Ste}

\def\solvref{BP,BH,HaA,ShP}

\begin{document}

\title{Simple and collective twisted symmetries}

\author{G. Gaeta\thanks{Research partially supported by MIUR-PRIN program
under project 2010-JJ4KPA} \\
{\it Dipartimento di Matematica, Universit\`a degli Studi di Milano,} \\
{\it via C. Saldini 50, 20133 Milano (Italy)} \\ {\tt
giuseppe.gaeta@unimi.it}}

\date{}

\maketitle

\begin{abstract} After the introduction of $\lambda$-symmetries by Muriel and Romero,
several other types of so called ``twisted symmetries'' have been
considered in the literature (their name refers to the fact they
are defined through a deformation of the familiar prolongation
operation); they are as useful as standard symmetries for what
concerns symmetry reduction of ODEs or determination of special
(invariant) solutions for PDEs and have thus attracted attention.
The geometrical relation of twisted symmetries to standard ones
has already been noted: for some type of twisted symmetries (in
particular, $\lambda$ and $\mu$-symmetries), this amounts to a
certain kind of gauge transformation.

In a previous review paper \cite{Gtwist} we have surveyed the
first part of the developments of this theory; in the present
paper we review recent developments. In particular, we provide a
unifying geometrical description of the different types of twisted
symmetries; this is based on the classical Frobenius reduction
applied to distribution generated by Lie-point (local) symmetries.

\end{abstract}

\section*{Introduction}

Symmetry analysis of differential equations is a classical topic
at least since the times of Sophus Lie, who indeed created what is
nowadays known as the theory of Lie groups as a tool to analyze
differential equations. In particular, symmetry properties are
used both to obtain (symmetry) reduction of ODEs and special
(symmetric) solutions to PDEs. The general standard scheme of Lie
group analysis of differential equations \cite{\symmref} goes with
considering the phase manifold $M = B \times U$, where $B$ is the
manifold of independent variables $x^i$ and $U$ that of dependent
variables $u^a$, and Lie-point transformations acting in this.
Once we know how the $x$ and the $u$ are transformed, we also know
how the derivatives (of the $u^a$ with respect to the $x^i$, of
any order) are transformed, so that the transformation considered
in $M$ is readily {\it prolonged} to a transformation in the Jet
bundle $J^n M$ (see Section \ref{sec:background} below for
detail). Having defined a transformation in the space of $x$, $u$,
and derivatives of any order -- in particular, up to the order of
the differential equation or system under study -- we can
ascertain if the transformation leaves the equation invariant, in
which case it qualifies as a {\it symmetry} of the latter, and can
be used for symmetry reduction or for determining special
invariant solutions. This scheme was generalized in several
directions as for the kind of transformations to be considered in
$M$ as a starting point for the whole procedure (e.g., Lie himself
considered contact transformations, in which the change of $x$,
$u$ depends on the first order derivatives as well; more generally
one can consider generalized maps depending on derivatives of
order $m$); but a firm point of the analysis is that once the map
acting in $M$ is given, the action on derivatives follows from
that on basic variables. It came thus as a surprise when, about
ten years ago, Muriel and Romero \cite{MuRom1,MuRom2} came out
with a different proposal: that is, consider standard Lie-point
transformation in $M$, but {\it deform the prolongation
operation}. The deformation they considered depends on a smooth
function $\la : J^1 M \to \R$, so that they christened these as
$\la$-prolongations and, when the prolonged vector field in $J^n
M$ leaves the equation invariant, {\it $\la$-symmetries}; standard
ones are recovered for $\la=0$, so that we have an extension of
standard Lie-point symmetries. What was more relevant, they showed
that such $\la$-symmetries were as useful as standard ones for the
symmetry reduction of ODEs (and systems thereof). As
$\la$-symmetries are more general than standard ones, this opens
the way at the possibility of applying symmetry techniques -- in
particular, symmetry reduction, and maybe a full solution -- to
equations with a lack of standard symmetries. This was indeed the
motivation behind the work of Muriel and Romero, and in their
first works they were immediately able to obtain concrete results
in this direction. The approach followed for the creation of
$\la$-symmetry was inherently analytic; but the theory of
symmetries of differential equations is deeply rooted in Geometry
(of Jet bundles), and a geometric understanding was needed. A
first step in this direction was obtained by Pucci and Saccomandi
\cite{PuS}, who noted that $\la$-prolonged vector field have a
peculiar characterization, being collinear to the standard
prolongation of some (in general, different) vector field in $M$.
This opened the way to a geometric understanding of the matter; it
turns out that the ODEs case is a degenerate one in this respect,
and hides several features, so that a full understanding is better
reached in the more general case of PDEs. In this framework the
smooth function $\la$ is replaced by a semi-basic one form $\mu =
\La_i \d x^i$ in $J^1 M$, the $\La_i$ being smooth matrix
functions (their dimension being the same as the dimension of
$U$), and one speaks accordingly of $\mu$-prolongations and {\it
$\mu$-symmetries} \cite{CGMor,GMor}. The $\La_i$ have to obey some
compatibility condition (see Sect.\ref{sec:PDE} below), but in the
ODE case there is only one $\La$, hence no compatibility. The
special case of $\mu$-symmetries for ODEs has been studied in
detail by Cicogna \cite{Cds1,Cds2,Cic13}; to emphasize its
intermediate character one speaks of $\La$-symmetries, and
sometimes of $\rho$-symmetries when dealing with the class of
$\La$-symmetries which effectively leads to a reduction of ODEs
(in this case the $\rho$ stands for ``reducing''). The collective
name for these deformed prolongations and symmetries is that of
{\it twisted} prolongations and symmetries. These developments,
and their geometrical aspects, were reviewed in detail in a
previous work by the present author \cite{Gtwist}, describing the
situation about five years ago and in particular the connection
with {\it gauge transformations}. In the present paper, we want to
review further and more recent  progress. This focuses in
particular on the possibility of considering deformed
prolongations not of a single vector field but of a set (in
involution) of vector fields, each of them entering in the
prolongation of the others. This induces naturally to focus on the
(Frobenius) {\it distributions} generated by the set of vector
fields, and this shift of focus allows to finally reach a sound
understanding of the geometrical aspects of twisted prolongations,
explaining the reason for the appearance of gauge transformations.

\medskip\noindent
{\bf Acknowledgements.} I thank the referees for useful remarks and suggestions. My research is partially supported by
MIUR-PRIN program under project 2010-JJ4KPA.

\section{Background material}
\label{sec:background}
\def\rs{\ref{sec:background}}

We start by recalling the usual symmetry reduction procedure
(under standard symmetries) for ODEs or systems thereof; this will
allow to fix notation and be useful for later reference. We refer
to \cite{Olv1} or also e.g.
\cite{EMS1,CGbook,Gaebook,KrV,Oli,Olv2,Ste} for details. On the
other hand, we assume the reader is familiar with the basic
notions in the theory of symmetry of differential equations (and
its use for solving these); see e.g. again \cite{\symmref}.

\subsection{General framework}
\label{sec:general}

We denote the system (which could reduce to a single equation) of
ODEs under study as $\De$, and assume for simplicity that it is in
standard form, i.e. is solved w.r.t. higher order derivatives. We
will write the independent variable as $x$ and the dependent ones
as $u^a$ ($a = 1,...,p$); derivatives of these will be denoted as
$u^a_x$, $u^a_{xx}$, and so on. We will also use the notation
$$ u^a_{(n)} \ := \ \frac{d^n u^a}{d x^n} \ . $$
We thus have  $\De$ written in the form \beq\label{eq:DE}
u^a_{(n)} \ = \ F^a (x,u,...,u_{(n-1)} ) \ . \eeq

The extended phase manifold will be $M = B \times U$, where $B =
\R$ (with $x \in B$) and $u = (u^1 , ... , u^p )$ takes values in
$U \subseteq \R^p$. We consider Lie-point vector fields in $M$;
these are written in coordinates as \beq\label{eq:X} X \ = \ \xi
(x,u) \ \frac{\pa}{\pa x} \ + \ \phi^a (x,u) \ \frac{\pa}{\pa u^a}
\ \equiv \ \xi \, \pa_x \ + \ \phi^a \, \pa_a \ . \eeq Here and in
the following, we use the Einstein sum convention.

As well known, \eqref{eq:DE} identifies a submanifold $S_\De$ in
the $n$-th order Jet space over $M$, $S \ss J^{(n)} M$; this is
called the {\it solution manifold} for $\De$. It is also well
known that $X$ induces a vector field $X^{(n)}$ in $J^{(n)} M$
(this is also known as the {\it prolongation} of $X$); then $X$ is
a Lie-point symmetry of $\De$ if and only if \beq\label{eq:sym}
X^{(n)} \ : \ S_\De \, \to \, \T S_\De \ . \eeq The same relation
can be expressed algebraically, considering $X$ and $X^{(n)}$ as
differential operators, as \beq\label{eq:symmalg} \[ X^{(n)} (\De)
\]_{S_\De} \ = \ 0 \ . \eeq

It is  convenient, for later reference, to recall how $X^{(n)}$ is
built from $X$. With the notation already introduced, we write $$
X^{(n)} \ = \ \xi \, \frac{\pa}{\pa x} \ + \ \psi^a_{(n)} \
\frac{\pa }{\pa u^a_{(n)}} \ ; $$ then the {\it prolongation
formula} is given in recursive terms as
\beq\label{eq:prolformODEs} \psi^a_{(n+1)} \ = \ D_x (\psi^a_{(n)}
) \ - \ u^a_{(n+1)} \ D_x \xi \ , \eeq where $\psi^a_{(0)} =
\phi^a$; here and below $D_x$ is the total derivative operator
(with respect to $x$), \beq D_x \ = \ \frac{\pa}{\pa x} \ + \
u^a_x \, \frac{\pa}{\pa u^a} \ + \ u^a_{xx} \, \frac{\pa}{\pa
u^a_x} \ + \ ... \ = \ \frac{\pa}{\pa x} \ + \ u^a_{(k+1)} \,
\frac{\pa}{\pa u^a_{(k)}} \ . \eeq

We also recall, again for later reference, that to the vector
field $X$ is associated its {\it evolutionary representative}
$X_v$; with $X$ as in \eqref{eq:X}, this is defined by
\beq\label{eq:Xv} X_v \ = \ (\phi^a \, - \, u^a_x \, \xi ) \
\frac{\pa}{\pa u^a} \ := \ Q^a \ \frac{\pa}{\pa u^a} \ . \eeq

As well known, $X^{(n)} : S_\De \to \T S_\De$ if and only if
$X_v^{(n)} : S_\De \to \T S_\De$\footnote{Note that while $X_v$ is
a generalized vector field on $M$, its prolongation $X_v^{(n)}$,
when restricted to $S_\De$, is a vector field on $S_\De \subset
J^{(n)} M$.}. In other words $X$ is a symmetry of $\De$ if and
only if $X_v$ is a symmetry for the same system $\De$ (see
Proposition 5.5 in \cite{Olv1}).

It is easy to check by explicit computations (e.g. in recursive
form) that
$$ Q^a_{(k)} \ = \ \psi^a_{(k)} \ - \ u^a_{(k+1)} \, \xi \ . $$
It follows from this that the prolongation of a vector field $X$
of the form \eqref{eq:X} and of its evolutionary representative
$X_v = Q^a \pa_a$ are related through \beq X_v^{(k)} \ = \ X^{(k)}
\ - \ \xi \, \pa_x \ - \ [u^a_{(k+1)} \, \xi ] \, \pa_a^k \ . \eeq

Finally, we say that $X$ is a {\it strong symmetry} for $\De$ if
$X^{(n)} (\De) = 0$; i.e., at difference with \eqref{eq:sym} and
\eqref{eq:symmalg}, we do not consider restriction to $S_\De$.

\medskip\noindent
{\bf Remark 1.} It may be proved that if $X$ is a (standard)
Lie-point symmetry for $\De$, then there is an equivalent equation
$\wt{\De}$ which admits $X$ as a strong symmetry \cite{CDW}.
Equivalence here is meant as admitting the same set of solutions;
in other words, we have $$ \De \ = \ e^{\mathcal{F}} \ \wt{\De} \
, $$ with $\mathcal{F} = \mathcal{F} (x;u,...,u_{(n)} )$. Such a
$\wt{\De}$ is necessarily written in terms of the differential
invariants for $X$ \cite{Olv1}, see below. \EOR
\bigskip

Let us now consider solutions to (a system of) differential
equations; a function $u = f(x)$, $u : B \to U$, is identified
with its graph, i.e. with a section $\ga_f$ of the bundle
$(M,\pi_0,B)$ with fiber $\pi_0^{-1} (x) \simeq U$. This is just
$$ \ga_f \ = \ \{ (x,u)  \in M \ : \ u  = f(x) \} \ . $$
The section is naturally lifted (or {\it prolonged}) to a section
$\ga_f^{(n)}$ of $(J^{(n)} M , \pi_n , B)$, i.e. of $J^{(n)} M$
seen as a bundle over $B$; here $\pi_n^{-1} (x) = (u,u_{(1)} , ...
, u_{(n)} )$.

The function $u = f(x)$ is a solution to the equation $\De$ of
order $n$ if and only if $\ga_f^{(n)} \subset S_\De$. Thus a
symmetry of $\De$ maps solutions into (generally, different)
solutions. (This can also be taken as the definition of
symmetries.)

We say that $u= f(x)$ is an {\it invariant solution} (under $X$)
to $\De$ if it is a solution and $X : \ga_f \to \T \ga_f$. This
condition is completely equivalent to requiring that $X^{(n)} :
\ga_f^{(n)} \to \T \ga_f^{(n)}$ \cite{\symmref}. Similarly, if we
consider a Lie algebra $\G$ of vector fields in $M$, we say that
$u = f(x)$ is a $\G$-invariant solution to $\De$ if it is a
solution and $X : \ga_f \to \T \ga_f$ for all $X \in \G$.

\medskip\noindent
{\bf Remark 2.} We stress that when requiring e.g. $X : \ga_f \to
\T \ga_f$, we are considering $\ga_f$ as a subset of $M$; if we
consider $\ga_f$ as a point in the space $\Sigma (M,\pi_0,B)$ of
smooth sections of $(M,\pi_0,B)$, and denote by $\Phi_X$ the flow
induced by $X$ in this space, we would rephrase this requirement
as $\Phi_X : \ga_f \to \ga_f$. \EOR

\subsection{Symmetry reduction of ODEs}
\label{sec:ODE}

Let us now recall the standard symmetry reduction procedure for
ODEs; this is based on {\it adapted coordinates}. For the sake of
simplicity, we will first discuss the case where $\De$ is a single
ODE of order $n$, $U \simeq \R$ and $M \simeq \R^2$. We assume
that $X$ is a Lie-point symmetry for $\De$, thus
\beq\label{eq:strsymm} \[ X^{(n)} (\De) \]_{S_\De} \ = \ 0 \ . \eeq
Now the key idea is to pass to new coordinates $(y,v)$ in $M$
(where $y$ should be thought as the independent variable) such
that in the new coordinates $X$ is written as \beq\label{eq:Xsa} X
\ = \ \frac{\pa }{\pa v} \ . \eeq (Alternatively, it may suffice
to get coordinates such that $X = \psi (y,v) (\pa /\pa v)$, with
$\psi$ nonzero; this requires a more detailed discussion.)

The point is that as symmetry is a tangency condition, see
\eqref{eq:sym}, and such a condition is independent of the
coordinate description, we are guaranteed that expressing $\De$ in
the new coordinates we still have \eqref{eq:sym}; to stress we are
expressing $\De$ in the new coordinates\footnote{We trust there is
no possible confusion between the equation as a geometrical object
(a submanifold in Jet space) and its coordinate expression.}, we
will also write it as $\^\De$.

On the other hand, \eqref{eq:Xsa} immediately entails that \beq
X^{(n)} \ = \ \frac{\pa }{\pa v} \ = \ X \ ; \eeq thus $\^\De$
admits $X$ as a symmetry if and only if it can be written with no
explicit dependence on $v$. In other words, the (new) coordinate
expression of our equation will be \beq \frac{\pa^n v}{\pa y^n} \
= \ G (y; v , v_{(1)}, ... , v_{(n-1)} ) \eeq with $G$ actually
independent of $v$,
$$ \pa G / \pa v \ = \ 0 \ . $$
We can thus rewrite $\^\De$ as \beq \frac{\pa^n v}{\pa y^n} \ = \
G (y; v_{(1)}, ... , v_{(n-1)} ) \ . \eeq

It suffices to pass to the new variable \beq\label{eq:w} w  \ := \
v_{(1)} \ = \ (\pa v / \pa y) \eeq to get an equation of order
$(n-1)$, i.e. to rewrite $\^\De$ as \beq\label{eq:DeR}
\frac{\pa^{n-1} w}{\pa y^{n-1}} \ = \ G (y; w , w_{(1)}, ... ,
w_{(n-2)} ) \ . \eeq

Note that if we have a solution to \eqref{eq:DeR}, this provides
solutions to the original equation $\De$: in fact, \eqref{eq:w}
yields $v$ just by a quadrature (which introduces an integration
constant), and this gives a solution to $\^\De$. Then we just have
to invert the change of variables $(x,u) \to (y,v)$ to get a
solution to $\De$.

In the case of a system, a Lie-point symmetry allows in this way
to reduce by one the order of one of the equations. In the case
where we have several Lie-point symmetries with a favorable
Lie-algebraic structure (e.g. they should form a solvable algebra)
we can correspondingly reduce several equations in the system, or
reduce the order of a single equation by several degrees. We refer
to \cite{\symmref} for further detail.

\medskip\noindent
{\bf Remark 3.} Note that the reduction approach can be improved
by considering so called {\it solvable structures}; the reader is
referred to \cite{\solvref} for detail on these (see also
\cite{MS,CFM1,CFM2} for recent results bearing connection to our topic).
\EOR

\subsection{Differential invariants and IBDP}
\label{sec:IBDP}

The discussion of the previous subsection is not entirely
satisfactory, in that it is based on coordinate representations.
In order to make it coordinate-free, we should remark that a
differential equation admitting a given vector field $X$ as a
Lie-point symmetry should be written in terms of the {\it
differential invariants} for $X$, possibly plus a nowhere
vanishing non-invariant term (see subsection \ref{sec:ODE}, in
particular Remark 1, above).

Thus -- up to the nowhere vanishing and thus irrelevant (in the
search for solutions) term, which we can write as $e^\F$ -- the
equation $\De$ is written as
$$ \De \ = \ \Phi [ \eta , \zeta_{(1)} , ... , \zeta_{(n)} ] \ , $$
where $\zeta_{(k)} : J^{(k)} M \to \R$ are the differential
invariants of order $k$, and $\eta : M \to \R$ invariants of order
zero (geometrical invariants).

It is essential to note that once we have differential invariants
of order one, those of higher order can be obtained simply by
derivation. In fact, we have the so called  ``invariant by
differentiation property'' (IBDP); this is well known and we give
it without proof (for a proof see e.g. \cite{Olv1}; see also our
proof of Lemma 5 below).\footnote{The relevance of IBDP for twisted
symmetries was recognized since the beginning; e.g. Muriel and
Romero emphasized it for $\la$-symmetries in \cite{MuRom3}.}

\medskip\noindent
{\bf Lemma 1 (IBDP).} {\it Let $\eta$ be a geometrical invariant
for $X$, and $\zeta_{(k)}$ a differential invariant of order $k$
for $X$. Then \beq \zeta_{(k+1)} \ := \ \frac{D_x \zeta_{(k)}
}{D_x \eta} \eeq is a differential invariant of order $k+1$ for
$X$.}

\subsection{Multiple symmetries}
\label{sec:multisym}

We have so far considered a single vector field; we could of
course have several Lie-point symmetries $X_i$, and in view of
\eqref{eq:sym} it is clear that the prolonged vector fields
$X_i^{(n)} = Y_i$ will form a Lie algebra. That is, if
$\mathcal{Y}_\De$ is the set of vector fields in $J^n M$ which
satisfy \eqref{eq:sym}, then $Y_i \in \mathcal{Y}_\De$, $Y_j \in
\mathcal{Y}_\De$ implies that $[Y_i , Y_j ] \in \mathcal{Y}_\De$
as well.

The possibility of using, at least in principles\footnote{In
practice, this will depend on the Lie algebraic structure of the
symmetry algebra.}, all the symmetry vector fields to obtain a
reduction of the $\De$ under study, and actually the very
possibility of speaking of a {\it symmetry algebra} for $\De$,
rely upon the following well known fact (this is Theorem 2.39
in \cite{Olv1}; see the proof in there):

\medskip\noindent
{\bf Lemma 2.} {\it Let $X_i$ be Lie-point vector fields on $M$,
and $X^{(n)}_i$ their prolongation on $J^n M$. The prolongation of
a commutator is the commutator of the prolongations, so the
$X_i^{(n)}$ satisfy the same Lie-algebraic relations as the $X_i$:
$$ [X_i^{(n)} , X_j^{(n)} ] \ = \ ([X_i,X_j])^{(n)} \ . $$}


\subsection{Invariant solutions for PDEs}
\label{sec:PDE}

The situation is quite different when we analyze PDEs. Albeit the
results mentioned above do hold also in this case, they are of no
great use since the dimension of Jet spaces $J^{(n)} M$ grows
combinatorially with $n$. Thus the IBDP only provides a subset of
the needed differential invariants to obtain a reduction. On the
other hand, symmetry analysis allows to obtain a reduced equation
which describes the symmetry-invariant solutions.

Discussing symmetries for PDEs requires to change slightly our
notation, renouncing to the simplifications made possible by
dealing only with the ODE case.

We now have $B \subseteq \R^q$ as the space for independent
variables  $x^i$ ($i=1,...,q$); we will now write a generic vector
field in $M$ as \beq\label{eq:Xgen} X \ = \ \xi^i (x,u) \
\frac{\pa}{\pa x^i} \ + \ \phi^a (x,u) \ \frac{\pa}{\pa u^a} \ = \
\xi^i \pa_i \ + \ \phi^a \pa_a \ . \eeq The partial derivatives of
the $u^a$ will be denoted as $u^a_J$, with $J = (j_1 , ... , j_q
)$ a multi-index of order $|J|= j_1 + ... + j_q$; here
$$ u^a_J \ := \ \frac{\pa^{|J|} u^a}{\pa (x^1)^{j_1} ... \pa (x^q)^{j_q} } \ . $$
We also write
$$ u^a_{J,i} \ := \ \frac{\pa u^a_J}{\pa x^i} \ = \ \pa_i u^a_J \ . $$

For several independent variables we have correspondingly several
total differential operators
$$ D_i \ = \ \frac{\pa}{\pa x^i} \ + \ u^a_{J,i} \ \frac{\pa}{\pa u^a_J}
\ = \ \pa_i \ + \ u^a_J \, \pa^J_a \ . $$
Note that here and in the following we use the lazy notation
$$ \pa^J_a  \ := \ \frac{\pa}{\pa u^a_J} \ . $$

The prolonged vector field $X^{(n)}$ will then be written as \beq
X^{(n)} \ = \ \xi^i \, \pa_i \ + \ \psi^a_J \, \pa_a^J \ , \eeq
where $\psi^a_0 = \psi^a$ and the other coefficients are given by
the standard prolongation formula; with our notation this is
\beq\label{eq:prolformPDEs} \psi^a_{J,i} \ = \ D_i \psi^a_J \ - \
u^a_{J,k} \, D_i \xi^k \ . \eeq

\medskip\noindent
{\bf Remark 4.} Note that the prolongation of a vector field $X$
of the general form \eqref{eq:Xgen} and of its evolutionary
representative $X_v = Q^a \pa_a$, where now $Q^a = \phi^a - u^a_i
\xi^i$, are related through $$ X_v^{(n)} \ = \ X^{(n)} \ - \ \xi^k
\, \pa_k \ - \ u^a_{J,k} \, \xi^k \, \pa_a^J \ . \eqno\odot $$
\bigskip

Given an algebra $\G$ of Lie-point vector fields on $M$,
$\G$-invariant solutions for $\De$ are solutions which are
invariant under all of $\G$. Thus one can determine them through a
symmetry reduction, which is not acting on the equation, but only
on the space of solutions with the desired symmetry properties. In
practice, one pass to a system of partially adapted coordinates
(some of them being invariant under $\G$, and residual ones), and
look for solutions which are expressed uniquely in terms of the
invariant ones.

This can be done by complementing the system $\De$ with a system
$\De_\G$ expressing the invariance condition; for an algebra with
generators $$ X_\a \ = \ \xi^i_\a \, \pa_i  \ + \ \phi^a_\a \,
\pa_a \ \ \ \ (\a = 1,...,r ) \ , $$ this is written as
$$ \phi^a \ - \ \xi^i \ \frac{\pa u^a}{\pa x^i} \ = \ 0  \ . $$

\medskip\noindent
{\bf Remark 5.} It should be noted that, albeit in general one
looks for $\G$-invariant solutions with $\G$ a symmetry
(sub)algebra for $\De$, there may exist solutions which are
invariant under more general vector fields (a trivial example is
often provided by the null solution); in this case one speaks of
``conditional symmetries'' \cite{LeWi} (see \cite{CGpart} for a
generalization) or of ``weak symmetries'' \cite{CK,PuSweak};
the notion of ``side conditions'' \cite{OR1,OR2} (see also \cite{CGWjlt1,CGWjlt2}) is also used in this context.
\EOR

\section{Twisted symmetries}
\label{sec:twist}

Twisted symmetries were first introduced by Muriel and Romero in
2001 \cite{MuRom1} in the form of $\la$-symmetries for scalar ODEs
and soon extended to a more general setting
\cite{MuRom2,MuRom3,MuRom4}; see also
\cite{MuRom4b,MuRom5,MuRom6,MuRom7,MuRom8,MRO} for further developments;
for a comprehensive review up to 2009 see \cite{Gtwist}.

Twisted symmetries are standard Lie-point vector fields in $M$
which are prolonged to vector fields in $J^{(n)} M$ through a
modified (``twisted'') prolongation operation, usually based on a
auxiliary object: a function $\la$ for $\la$-symmetries, a
one-form $\mu$ for $\mu$-symmetries, a matrix $\La$ or $\s$ for
$\La$- or $\s$-symmetries. In this way we get a twisted
prolongation $X^{(n)}_\theta = Y$; we say that $X$ is a twisted
symmetry for a (partial or ordinary) differential equation $\De$
if this twisted prolongation satisfies the analogue of
\eqref{eq:sym}, i.e. if \beq\label{eq:symtwist} Y \ : \ S_\De \
\to \ \T S_\De \ . \eeq Obviously this geometrical condition is
also written in algebraic terms as \beq \[ Y (\De ) \]_{S_\De} \ =
\ 0 \ . \eeq

We will divide twisted prolongations, and hence twisted
symmetries, in two classes:

\begin{itemize}
 \item[(A)] Those for which there is a modified prolongation
operation for the single vector field; we call these ``simple
twisted symmetries''.
 \item[(B)] Those for which we consider
several vector fields (generating an involution system) at once,
and the modified prolongation operation acts on each vector field
depending also on other vector fields; we call these ``collective
twisted symmetries''.
\end{itemize}

Obviously one could consider the first class as a special case of
the second one; however we feel that it is convenient to treat
them separately. This is for several reasons: first of all because
this in this way our discussion will be more clear; and then also
to follow the historical development and to the benefit of the
reader who is interested only in ``classical'' twisted symmetries
(those of the first type). Moreover, as some geometrical
understanding of type (A) is provided in the literature, in terms
of gauge theories, it is appropriate to devote a special
discussion of how this fits into the framework we are proposing
here. We will discuss the gauge properties of different types of
twisted symmetries, and later on how these are natural in view
of the picture which emerges from our discussion.

\section{Simple twisted symmetries}

\subsection{Simple twisted symmetries: $\la$-symmetries}
\label{sec:lasymm}

The prototype of simple twisted symmetries is given by the
$\la$-symmetries introduced by Muriel and Romero
\cite{MuRom1,MuRom2, MuRom3,MuRom4,MuRom4b,MuRom5,MuRom6}; these were then
generalized to $\mu$-symmetries \cite{CGMor,GMor} (which deal also
with PDEs) and to $\La$- and $\rho$-symmetries \cite{Cds1,Cds2}.

In $\la$-symmetries, the prolongation operation is modified via an
auxiliary smooth real function $\la : J^{(1)} M \to \R$; then
\eqref{eq:prolformODEs} is replaced by \beq\label{eq:lambdaprol}
\psi^a_{(k+1)} \ = \ (D_x + \la ) \psi^a_{(k)} \ - \ u^a_{(k+1)}
\, (D_x + \la ) \xi \ . \eeq We say then that
$$ Y \ = \ X^{(n)}_\la \ = \ \xi \, \frac{\pa}{\pa x} \ + \
\psi^a_k \, \frac{\pa}{\pa u^a_k} $$ is the {\it
$\la$-prolongation} of $X$.

If the equation $\De$ is invariant under the $\la$-prolongation of
$X$, i.e. if
$$ Y \ : \ S_\De \ \to \ \T S_\De \ , $$ we say that $\De$ admits
$X$ as a {\it $\la$-symmetry}.

\medskip\noindent
{\bf Remark 6.} Note that here not only the modification of the
prolongation acts autonomously on each vector field, but actually
each component (in the sense of the vector structure in $U$ and on
the associate ``derivative'' spaces $U_{(k)} \subset J^{(n)} M$
for $k \le n$) is transformed autonomously. It is also quite clear
from \eqref{eq:lambdaprol} that we can interpret this as the
replacement of the standard total derivative $D_x$ with a sort of
``total covariant derivative'' \beq \nabla_x \ := \ D_x \ + \ \la
\ ; \eeq adopting this point of view, $\la$ is associated to a
connection in a real line bundle over $B$, and this acts on all
line bundles associated to different components of $u^a$ (a more
precise geometrical discussion will be provided below). \EOR

\medskip\noindent
{\bf Remark 7.} It is also natural, in view of
\eqref{eq:lambdaprol}, to consider the case where we do not have
identical connections on these different line bundles: this
corresponds to considering different functions $\la_a$ for
different components $\psi^a$. In other words we could have (with
no sum on $a$) \beq\label{eq:LAmbdaprol} \psi^a_{(k+1)} \ = \ (D_x
+ \la_a ) \psi^a_{(k)} \ - \ u^a_{(k+1)} \, (D_x + \la_a ) \xi \ .
\eeq This was considered in \cite{MuRom4}, and is also a special
case of Cicogna's $\La$-symmetries \cite{Cds1,Cds2,Cic13}, see
Sect.\ref{sec:musymm}. \EOR
\bigskip

If we compare $\la$-prolongations and symmetries with standard
prolongations and symmetries with our mind turned towards symmetry
reduction of ODEs, even of a single one, a striking difference is
immediately apparent: while for standard prolongations the vector
field $X = (\pa / \pa u)$ has a trivial prolongation, i.e.
$X^{(n)} = X$ -- see the discussion in section \ref{sec:ODE} --
now the same vector field has in general a nontrivial
prolongation, due to the function $\la$. This means in particular
that if we pass to symmetry-adapted coordinates, so that $X$ has
the above form, having an equation $\De$ which admits $X$ as a
$\la$-symmetry (with a nontrivial $\la$) does not entail that
$\De$ is independent of $u$. It is thus evident that our
discussion of section \ref{sec:ODE}, based on coordinates, does
not (at least directly) apply to symmetry reduction under
$\la$-symmetries.\footnote{Actually, there seem to be no {\it
apriori} reason for a symmetry reduction under $\la$-symmetries to
be possible. That this is the case, and actually that a kind of
symmetry reduction is possible considering deformed prolongation
operations, was first proved by analytical computation by Muriel
and Romero.}

On the other hand, it turns out that $\la$-prolongations possess
the IBDP \cite{MuRom3}; this shows that symmetry reduction is
actually possible under $\la$-prolonged vector fields. As this is
at the basis of the application of $\la$-symmetries, we will state
this as a Lemma. This result is well known and, as for Lemma 1 above, we will not provide a proof (see e.g. the original proof by Muriel and Romero \cite{MuRom3} or the discussion in \cite{Gtwist}; see also our proof of Lemma 5 below).

\medskip\noindent
{\bf Lemma 3.} {\it The IBDP holds for $\la$-prolonged vector fields.}
\bigskip

The discussion in \cite{MuRom3} provides a direct proof of the IBDP, based on
analytical computations. It is  also possible to proceed in a
different way. In fact, it turns out that $\la$-prolonged vector
fields are locally equivalent to ordinarily prolonged ones via a
gauge transformation, and one can infer the IBDP for the
$\la$-prolonged field from the fact it holds for the equivalent
ordinarily-prolonged one. We will now discuss this point of view,
starting with a discussion of the gauge equivalence
\cite{CGMor,GMor} (see also \cite{GJPAgau} for detail).

We stress beforehand that the gauge equivalence is in general only
{\it local} (and not global); thus twisted symmetries are in fact more general than standard ones, even considering gauge equivalence classes.

Moreover, this local gauge equivalence is in general realized through non-local functions, i.e. functions expressed as integrals of standard smooth functions. See in this respect the discussion in Remark 10 below.\footnote{As these integrals appear as arguments of an exponential function, and to avoid confusion with locality in the sense just mentioned above, we will often speak of ``exponential type'' functions or ``exponential vector fields''; this follows the notation suggested by Muriel and Romero \cite{MuRom1,MuRom5}, who in turn refer to Olver \cite{Olv1} for the original definition.}

\medskip\noindent
{\bf Lemma 4.} {\it Let us consider the Lie-point vector fields
$X$ and $\wt{X}$ on $M$, with $X$ as in \eqref{eq:X} and $\wt{X} =
\b X$, with $\b : M \to \R$ a nowhere zero smooth ($C^\infty$)
function on $M$. Then
$$ \b \ (X^{(n)}_\la ) \ = \ \wt{X}^{(n)} \ , $$
with $\la = (D_x \b) \b^{-1}$.}

\medskip\noindent
{\bf Proof.} Our assumption means that
$$ \wt{X} \ = \ \wt{\xi} \, \pa_x \ + \ \wt{\phi}^a \, \pa_a \ ; \ \
\wt{\xi} = \b \xi \ , \ \ \wt{\phi}^a = \b \phi^a \ . $$
Let us now consider the first prolongations of these; for $\wt{X}^{(1)}$
we have, according to the standard prolongation formula,
\begin{eqnarray*} \wt{\psi}^a_{(1)} &=& D_x \wt{\phi}^a \ - \
u^a_x \, D_x \wt{\xi} \ = \ D_x (\b \phi^a) \ - \ u^a_x \, D_x ( \b \xi ) \\
&=& \b \ \( D_x \phi^a \, - \, u^a_x \, D_x \xi \) \ + \ (D_x \b) \ \( \phi^a \, - \, u^a_x \, \xi \) \\
&=& \b \ \psi^a_{(1)} \ + \ (D_x \b) \ ( \phi^a \, - \, u^a_x \,
\xi ) \\ &=& \b \ \[ \psi^a_{(1)} \ + \ \b^{-1} \, (D_x \b)
 \ \( \phi^a \, - \, u^a_x \, \xi \) \] \ . \end{eqnarray*}
It is easy to check that this holds for prolongations in $J^{(n)}
M$, for whatever $n$. In fact, suppose it holds for $k$; then
proceeding exactly as above we have
\begin{eqnarray*} \wt{\psi}^a_{(k+1)} &=& D_x \wt{\psi}^a_{(k)} \ - \ u^a_{(k+1)} \, D_x \wt{\xi} \ = \
D_x (\b \psi^a_{(k)}) \ - \  u^a_{(k+1)} \, D_x ( \b \xi ) \\
&=& \b \ \[ \psi^a_{(k+1)} \ + \ \b^{-1} \, (D_x \b) \ \(
\psi^a_{(k)} \, - \, u^a_{(k+1)} \, \xi \) \] \ . \end{eqnarray*}
As we have just checked the required property holds for $k=0$, the
proof is completed by recursion up to any desired degree. \EOP

\bigskip\noindent
{\bf Remark 8.} We have thus seen that the multiplication by $\b$
connects $X$ and $\wt{X}$ at the level of vector fields in $M$,
while at the level of vector fields in $J^{(n)} M$ it connects the
$\la$-prolongation of $X$ (with $\la = (D_x \b) \b^{-1}$) and the
standard prolongation of $\wt{X}$. As by assumption $\b \not= 0$
in all points of $M$, the vector fields $X$ and $\wt{X}$ are
collinear, and we have seen that the $\la$-prolongation of a
vector field $X$ is collinear (through a factor $\b$) to the
standard prolongation of a collinear (through the {\it same}
factor $\b$) vector field $\wt{X}$. We can summarize these
relations in the form of a commutative diagram: \beq \matrix{ X &
\mapright{\b} & \wt{X} \cr
 & & \cr
\mapdown{\la-prol} & & \mapdown{prol} \cr
 & & \cr
\ X^{(n)}_\la & \mapright{\b} & \ \wt{X}^{(n)}_0 \cr} \eeq The
functions $\la$ and $\b$ are related by
\beq\label{eq:labeta} \la \ = \ (D_x \b) \ \b^{-1} \ = \ D_x ( \log \b ) \ ; \eeq we can also write
\beq\label{eq:labeta2} \b \ = \ \exp \[ \int \la \, d x \] \ . \eeq
Note also that as $\b : M \to \R$, it results $\la : J^1 M \to
\R$; the $C^\infty$ smoothness of $\b$ entails $C^\infty$
smoothness of $\la$. \EOR

\medskip\noindent
{\bf Remark 9.} The effect of multiplication of a $\la$-prolonged vector field $X$ by a smooth function $f$ has been considered in \cite{MuRom1}; in particular, Lemma 5.1 in there shows that in this way one can obtain a new $\la$-prolonged vector field with a new function $\overline{\la} = \la - [X(f)/f]$; choosing $X(f)/f = \la$ one obtains $\overline{\la} = 0$. \EOR
\bigskip

\medskip\noindent
{\bf Remark 10.} The functions $\la (x,u,u_x)$ and $\b (x,u)$ are related through \eqref{eq:labeta}. If $\b$ is given, this straightforwardly defines $\la$. On the other hand, if $\la$ is given and we want to determine $\b$, the solution is given by \eqref{eq:labeta2}, but here it is in general impossible to write $\b$ as a local function, and it can only be expressed as an integral; we also say that it is a \emph{non-local} function (note no confusion should be made with locality in the sense of the domain of definition of a function). If $X = \xi \pa_x + \phi^a \pa_a$, we will get
$$ \wt{X} \ = \ e^{\int \la d x} [\xi \pa + \phi^a \pa_a ] \ , $$ i.e. an ``exponential vector field'' in the sense of Olver \cite{Olv1}, see also \cite{GoL,MuRom1,MuRom5}.\footnote{Note also that the primitive of a function $\la$, hence $\b$, is not uniquely defined; however adding a constant term to $\b$ would give a global gauge transformation, not relevant here.} \EOR

\subsection{Simple twisted symmetries: $\mu$-symmetries}
\label{sec:musymm}

A substantial generalization of $\la$-symmetries concerns the
possibility of defining twisted prolongation operators also for
PDEs\footnote{The name ``twisted'' gets its justification in this
framework, as we will see in a moment.}; discussing this frame
requires to use the general notation (see subsection
\ref{sec:PDE}).

Now the role of the function $\la$ is taken by a semibasic
one-form $\mu \in \Lambda^1 (J^1 M)$, which we write as \beq \mu \
= \ \La_i \ \d x^i \ ; \eeq the $\La_i$ are matrix functions, $\La
: J^1 M \to \mathtt{Mat} (\R,q)$.  The form $\mu$ should be such
that $\d \mu$ belongs to the Cartan ideal $\mathcal{C}$ \cite{God,Sha,Stern}
generated by the contact forms in $J^n M$, i.e. \beq
\label{eq:dmucart} \d \mu \in \mathcal{C} \ . \eeq This amounts to
the requirement that $\mu$ should satisfy the horizontal
Maurer-Cartan equation \beq \label{eq:hMC} \ D \mu \ + \
\frac{1}{2} \ [\mu , \mu] \ = \ 0 \ . \eeq If seen in terms of the
matrices $\La_i$, this is the compatibility condition (here the
commutator is the standard commutator between matrices) \beq
\label{eq:LAmu} D_i \, \La_j \ - \ D_j \, \La_i \ + \ [\La_i ,
\La_j ] \ = \ 0 \ . \eeq We note that defining the ``total
covariant derivatives'' \beq \nabla_i \ := \ D_i \ + \ \La_i \ ,
\eeq the previous formulas can be rewritten as the zero-curvature
condition \beq \label{eq:ZCC} [ \nabla_i \, , \, \nabla_j ] \ = \
0 \ . \eeq

The condition \eqref{eq:hMC} -- or equivalently \eqref{eq:LAmu},
or \eqref{eq:ZCC} -- guarantee that $\mu$ is  horizontally
closed\footnote{The appearance of total (hence horizontal)
derivatives is related to taking into account the contact
structure in $J^n M$; see \cite{CGMor,GMor} for details.}: \beq D
\mu \ := \  (D_i \La_j) \ \d x^i \wedge \d x^j \ = \ 0 \ . \eeq
This in turn is nothing else than a restatement of
\eqref{eq:dmucart}.

We now define recursively the $\mu$-prolongation of the vector
field \eqref{eq:Xgen} as \beq Y \ = \ X^{(n)}_\mu \ = \ \xi^i \,
\frac{\pa}{\pa x^i} \ + \ \psi^a_{(J)} \, \frac{\pa }{\pa
u^a_{(J)} } \ = \ \xi^i \pa_i \ + \ \psi^a_{(J)} \pa_a^J \ , \eeq
with $\psi^a_{(0)} = \phi^a$ and \beq \label{eq:muprol}
\psi^a_{(J,i)} \ = \ (\de^a_{\ b} \, D_i \ + \ (\La_i)^a_{\ b} )
\, \phi^b \ - \ u^b_{J,k} \, (\de^a_{\ b} \, D_i \ + \
(\La_i)^a_{\, b} ) \, \xi^k \ . \eeq

\medskip\noindent
{\bf Remark 11.} It should be noted that $\mu$-symmetries apply
both for (systems of) ODEs and (systems of) PDEs; in the first
case we actually have a single matrix $\La$ (and no compatibility
condition, so that this is in several sense a degenerate case),
and one also uses the name of {\it $\La$-symmetries} \cite{Cds1,Cds2}.
These will be specially relevant in the following, when considering
reduction of systems of ODEs. \EOR
\bigskip

We will start by discussing the IBDP for $\mu$-symmetries. Here it
happens that the IBDP does in general {\it not} hold for
$\mu$-prolonged vector fields. If the matrices $\La_i$ are
multiples of the identity, then the IBDP property holds; the
proof of this fact reduces to a variant of the proofs of Lemma 1 and
Lemma 3.

\medskip\noindent
{\bf Lemma 5.} {\it Let $\mu = \La_i \d x^i$, the matrices $\La_i$
being multiple of the identity, \beq\label{eq:L5} (\La_i)^a_{\ b}
\ = \ \la_i \ \de^a_{\ b} \eeq with $\la_i : J^1 M \to \R$ scalar
$C^\infty$ functions. The IBDP holds for such $\mu$-prolonged
vector fields.}

\medskip\noindent
{\bf Proof.} We will proceed as in the standard proof of Lemma 1 (see \cite{Olv1}); we will however use the more general notation (so to include also the case of PDEs -- albeit, as mentioned above, the IBDP property is
not so useful in this context). We thus have
\begin{eqnarray*}
D_i &=& \pa_i \ + \ u^a_{(J,i)} \, \pa_a^J \ , \\
Y   &=& \xi^i \, \pa_i \ + \ \psi^a_{(J)} \, \pa_a^J \ ;
\end{eqnarray*} it follows that
$$ [Y,D_i] \ = \ \[ \psi^a_{(J,i)} \ - \ (D_i \psi^a_{(J)} ) \] \, \pa_a^J \ - \ (D_i \xi^k) \, \pa_k \ . $$
If $Y$ is a $\mu$-prolongation, it follows from \eqref{eq:muprol}
that (adding and subtracting a factor proportional to $\xi^k
\pa_k$ through a function $\rho_i$) this reads
\begin{eqnarray*} [Y,D_i] &=& \[ (\La_i)^a_{\ b} \( \psi^b_J - u^b_{J,k} \xi^k \) \ - \ u^a_{(J,k)} (D_i \xi^k) \] \ \pa_a^J \ - \ (D_i \xi^k) \, \pa_k \\
&=& \[ (\La_i)^a_{\ b} \( \psi^b_J - u^b_{J,k} \xi^k \) \] \ \pa_a^J \ - \ (D_i \xi^k) \, \[ \pa_k \ + \ u^a_{(J,k)} \, \pa_a^J \] \\
&=& \[ (\La_i)^a_{\ b} \( \psi^b_J - u^b_{J,k} \xi^k \) \] \ \pa_a^J \ - \ (D_i \xi^k) \, D_k \\
&=& \[ \rho_i \, \xi^k \, \pa_k \ + \ (\La_i)^a_{\ b} \, \psi^b_J
\, \pa_a^J \] \ - \ \xi^k \ \[ \rho_i \, \pa_k \ + \ (\La_i)^a_{\
b} \, u^b_{(J,k)} \, \pa_a^J \] \\ & & \  - \ (D_i \xi^k) \, D_k \
; \end{eqnarray*} note that here the functions $\rho_i$ are
arbitrary ones. If in addition the matrices $\La_i$ are multiples
of the identity, see \eqref{eq:L5}, this reads
$$ [Y,D_i] \ = \ \[ \rho_i \, \xi^k \, \pa_k \ + \ \la_i \, \psi^a_J \, \pa_a^J \] \ - \ \xi^k \ \[ \rho_i \, \pa_k \ + \ \la_i \, u^a_{(J,k)} \, \pa_a^J \] \ - \ (D_i \xi^k) \, D_k \ . $$
Finally, we make use of our freedom in choosing the functions
$\rho_i$ to set $\rho_i = \la_i$, and obtain
\begin{eqnarray}
[Y,D_i] &=& \la_i \ \[ \xi^k \, \pa_k \ + \ \psi^a_J \, \pa_a^J \] \ - \ \la_i \, \xi^k \ \[ \pa_k \ + \ u^a_{(J,k)} \, \pa_a^J \] \ - \ (D_i \xi^k) \, D_k \nonumber \\
&=& \la_i \ Y \ - \ [ \la_i \, \xi^k \ + \ (D_i \xi^k) ] \, D_k \
. \label{eq:LAcomm} \end{eqnarray}

Now let us consider the IBDP; we want to prove that if $\eta$ and
$\zeta_{(m)}$ are differential invariants for $Y$, so is
$\zeta_{(m+1)} := (D_i \zeta_{(m)})/(D_i \eta)$, for any $i =
1,...,p$. We do of course have
$$ Y \[ \frac{D_i \zeta_{(m)} }{D_i \eta} \] \ = \ \frac{Y(D_i \zeta_{(m)}) \cdot (D_i \eta) \ - \ (D_i \zeta_{(m)}) \cdot Y (D_i \eta)}{(D_i \eta)^2} \ , $$ so that we have to prove that
$$ Y(D_i \zeta_{(m)}) \cdot (D_i \eta) \ = \ (D_i \zeta_{(m)}) \cdot Y (D_i \eta) \ . $$ This can be rewritten as
$$ [Y,D_i] (\zeta_{(m)}) \cdot (D_i \eta) \ + \ D_i [ Y(\zeta_{(m)}) ] \ = \ (D_i \zeta_{(m)}) \cdot [Y,D_i] (\eta) \ + \ D_i [ Y(\eta) ] \ ; $$
since by assumption $Y(\zeta_{(m)} ) = 0 = \Y (\eta)$, this
reduces in turn to
$$ [Y,D_i] (\zeta_{(m)}) \cdot (D_i \eta) \ = \ (D_i \zeta_{(m)}) \cdot [Y,D_i] (\eta) \ . $$
We can now make use of \eqref{eq:LAcomm}, and rewrite this as
\begin{eqnarray*} & &  \la_i \ Y (\zeta_{(m)}) \cdot (D_i \eta) \ - \ (\la_i \xi^k \, + \, D_i \xi^k) \, D_k (\zeta_{(m)}) \cdot (D_i \eta) \\ & & \ = \
\la_i \ (D_i \zeta_{(m)}) \cdot Y (\eta) \ - \  D_i (\zeta_{(m)})
\cdot (\la_i \xi^k \, + \, D_i \xi^k) \, (D_k \eta) \ ;
\end{eqnarray*}  recalling again $Y(\zeta_{(m)} ) = 0 = \Y
(\eta)$, we are reduced to the identity
$$ (\la_i \xi^k \, + \, D_i \xi^k) \cdot D_k (\zeta_{(m)}) \cdot (D_i \eta) \ = \
D_i (\zeta_{(m)}) \cdot (\la_i \xi^k \, + \, D_i \xi^k) \cdot (D_k
\eta) \ . $$ This is trivially true, and the proof is completed.
\EOP

\medskip\noindent
{\bf Remark 12.} This proof also provides a proof of Lemma 1 and Lemma 3; in fact Lemma 1 is obtained setting all the $\La_i$ (i.e. the $\la_i$) to zero, and Lemma 3 restricting to the one-dimensional case.

\medskip\noindent
{\bf Remark 13.} The $\mu$-prolonged vector fields with $\mu$ as
in \eqref{eq:L5} are also called $\La$-prolonged (to emphasize the
similarity with standard $\la$-prolongations). Note that, as
immediately apparent from the proof (and as can be checked by
simple examples) the IBDP does {\it not} hold for general
$\mu$-prolonged vector fields, i.e. when \eqref{eq:L5} is not
satisfied.

Note also that the proof of IBDP for $\La$-prolongations seems to
appear in the literature only by considering the evolutionary
representative of the vector field $X$ and its $\mu$-prolonged
corresponding vector field; in this respect, see also
Sect.\ref{sec:evolutionary} below. \EOR
\bigskip

It turns out that $\mu$-prolonged evolutionary vector fields are
locally equivalent to ordinarily prolonged ones via a gauge
transformation (and one can infer the IBDP for the $\La$-prolonged
field from the fact it holds for the equivalent
ordinarily-prolonged one). We will now discuss this point of view,
starting with a discussion of the gauge equivalence (see also
\cite{CGMor,GJPAgau} for detail); we stress this only holds for
vector fields in evolutionary form, see also Sect.\ref{sec:evolutionary}.

\medskip\noindent
{\bf Lemma 6.} {\it Let us consider the evolutionary vector fields
$X$ and $\wt{X}$ on $M$, with $$ X \ = \ Q^a \, \pa_a \ , \ \
\wt{X} \ = \ (A^a_{\ b} \, Q^b ) \ \pa_a \ , $$ with $A : M \to
\mathtt{Mat} (\R,q)$ a nowhere zero smooth ($C^\infty$) matrix
function on $M$. Then \beq\label{eq:L6} A \ (X^{(n)}_\mu ) \ = \
\wt{X}^{(n)} \ , \eeq with $\mu = (D A) A^{-1}$.}

\medskip\noindent
{\bf Proof.} First of all we note that \eqref{eq:L6} should be
meant as a shorthand notation for
$$ X^{(n)}_\mu \ = \ Q^a_J \, \pa_a^J \ , \ \
\wt{X}^{(n)} \ = \ \wt{Q}^a_J \, \pa_a^J $$ with $Q$ and $\wt{Q}$
related through
$$ A^a_{\ b} \ Q^a_J \ = \ \wt{Q}^a_J \ ; $$
and of course $Q^a_J$ obeying the $\mu$-prolongation formula and
$\wt{Q}^a_J$ the standard one, so that
$$ Q^a_{J,i} \ = \ D_i \, Q^a_J \ + \ (\La_i)^a_{\ b} \, Q^b_J \ , \ \
\wt{Q}^a_{J,i} \ = \ D_i \, \wt{Q}^q_J \ . $$ Let us write, for
all $J$,
$$ Q^a_J \ = \ A^a_{\ b} \ P^b_J \ , $$
with $A$ a $C^\infty$ and nowhere singular matrix function on $M$.
Then the $\mu$-pro\-lon\-ga\-tion formula requires
\begin{eqnarray*}
Q^a_{J,i} &=& D_i (Q^a_J ) \ + \ (\La_i)^a_{\ b} \ Q^b_J \\
&=& D_i (A^a_{\ b} \, P^b_J ) \ + \ (\La_i)^a_{\ b} \ A^b_{\ c} \, P^c_J  \\
&=& D_i (A^a_{\ b}) \, P^b_J \ + \ A^a_{\ b} \, D_i (P^b_J ) \ + \
(\La_i \, A)^a_{\ b} \, P^b_J  \ . \end{eqnarray*} On the other
hand, we know that $$ Q^a_{J,i} \ = \ A^a_{\ b} \ P^b_{J,i} \ ; $$
comparing these two formulas, we get
$$ P^a_{J,i} \ = \ D_i \, P^a_J \ + \ [A^{-1} \, D_i (A )]^a_{\ b} \ P^b_J \ + \
(A^{-1} \, \La_i \, A )^a_{\ b} \, P^b_J \ . $$ We conclude that
the $P^a_J$ satisfy the standard prolongation formula provided $A$
satisfies
$$ A^{-1} \ (D_i A) \ + \ A^{-1} \, \La_i \, A \ = \ 0 \ ; $$ equivalently, provided
$$ (D_i A) \ A^{-1} \ = \ \La_i \ . $$ As $D A = (D_i A) \d x^i$, the proof is completed. \EOP

\bigskip\noindent
{\bf Remark 14.} We have thus seen that the action of $A$ connects
$X$ and $\wt{X}$ at the level of (evolutionary) vector fields in
$M$, while at the level of vector fields in $J^{(n)} M$ it
connects the $\mu$-prolongation of $X$ (with $\mu = (D A) A^{-1}$)
and the standard prolongation of $\wt{X}$. We can summarize these
relations in the form of a commutative diagram: \beq \label{diag:mu}
\matrix{ X &
\mapright{A} & \wt{X} \cr
 & & \cr
\mapdown{\mu-prol} & & \mapdown{prol} \cr
 & & \cr
\ X^{(n)}_\la & \mapright{A} & \ \wt{X}^{(n)}_0 \cr} \eeq

The matrices $\La_i$ defining $\mu$ and $A$ are related by
\beq\label{eq:ALA1} \La_i \ = \ (D_i A) \ A^{-1} \ = \ D_i ( \log A ) \ ; \eeq we can also write
\beq\label{eq:ALA2} A \ = \ \exp \[ \int \La_i \, d x^i \] \ . \eeq
Note also that $\La_i : J^1 M \to \mathtt{Mat}(\R,q)$; the
$C^\infty$ smoothness of $A$ entails $C^\infty$ smoothness of the
$\La_i$. \EOR

\medskip\noindent
{\bf Remark 15.} In the case $\La_i = \la_i I$, the equivalence is through a
simple rescaling of the vector fields. \EOR
\bigskip

\medskip\noindent
{\bf Remark 16.} The discussion of Remark 10 also applies here. That is, the relation between $A$ and the $\La_i$ is described by \eqref{eq:ALA1}, but this can be read in two ways: if $A$ is given, we immediately obtain the $\La_i$; on the other hand, if the $\La_i$ are assigned and we look for the corresponding $A$, we obtain \eqref{eq:ALA2}, but it is in general impossible to express $A$ as a local function, and we are again led to consider exponential type vector fields \cite{MuRom1,MuRom5,Olv1}. \EOR
\bigskip

When we look for invariant solutions to PDEs, we focus on the
invariant sections of $(J^n M,\pi_n,B)$. It happens that the
invariant sets for the standard prolongation of a vector field and
for its $\mu$-prolongation do coincide. In fact, we have more
precisely the following results.

\medskip\noindent
{\bf Lemma 7.} {\it Let $X$ be an evolutionary vector field of the
form $X = Q^a \pa_a$. Denote by $X^{(n)}_\mu$ its standard
prolongation with $\mu = \La_i \d x^i$, and by $X^{(n)} =
X^{(n)}_0$ its ordinary prolongation; write
$$ X^{(n)}_\mu \ = \ \Psi^a_J \, \pa_a^J \ ; \ \ X^{(n)} \ = \ \Phi^a_J \, \pa_a^J \ . $$
Then the coefficients $\Psi^a_J$ (obeying the $\mu$-prolongation
formula) and $\Phi^a_J$ (obeying the standard prolongation
formula) are related by
$$ \Psi^a_J \ = \ \Phi^a_J \ + \ F^a_J \ ; $$
the difference term satisfies $F^a_0 = 0$ and obeys the recursion
relation
$$ F^a_{J,i} \ = \ [\de^a_b \, D_i \ + \ (\La_i)^a_{\ b} ] \,
F^b_J \ + \ (\La_i)^a_{\ b} \, (D_J Q^b) \ . $$}

\medskip\noindent
{\bf Proof.} This is Theorem 6 of \cite{GMor}; see there for the
detailed proof (which amounts to a rather involved explicit computation). \EOP

\medskip\noindent
{\bf Remark 17.} The $X$-invariant sections $\ga$ of $(M,\pi_0,B)$
are characterized by the vanishing of $Q$ on them (or more
precisely on their first prolongation $\ga^{(1)}$). It follows
from Lemma 7 that on the subset of sections with $Q=0$ we have
$F^a_J=0$ at all orders (recall $F^a_0 = 0$). Thus we conclude
that standard and $\mu$-prolongations of evolutionary vector
fields do coincide on the set of invariant sections, hence on
invariant functions. \EOR

\medskip\noindent
{\bf Remark 18.} As recalled in Remark 5 above, a particularly effective generalization of the standard symmetry approach to determining invariant solutions to PDEs is to consider ``weak'' or ``conditional'' symmetries \cite{CK,LeWi,OR1,OR2,PuSweak}. The idea is to look for transformations which are not proper symmetries of the PDEs under study (thus do not in generalmap solutions into solutions) but such that the PDEs admit solutions which are invariant under these transformations. One could wonder if an extension of this concept to the realm of twisted symmetries -- i.e. considering weak or conditional twisted symmetries -- is possible and effective. The previous Lemma and Remark show that this is possible but not effective, in that on the set of invariant sections we would reduce to consider standard prolongations and hence standard weak or conditional symmetries. The situation is different when considering so called ``partial symmetries'': these are transformations which map a subset of solution into other solutions of the same subset, and in the non trivial case the solutions will be mapped into different solutions. In this case the extension of the approach to twisted symmetries can give nontrivial results, as discussed in \cite{CGWjlt2}. \EOR

\medskip\noindent
{\bf Remark 19.} Here we are not discussing the relation of $\mu$-prolongations
and symmetries with $\mu$-related deformed exterior and Lie derivatives; this
point of view was advocated by Morando \cite{Mordef}.

\subsection{$\mu$-symmetries without evolutionary representatives}
\label{sec:evolutionary}

We have obtained our result about gauge equivalence of
$\mu$-prolonged vector fields with standardly prolonged ones at
the price of dealing exclusively with vertical vector fields; if
the vector fields we wish to consider are not vertical, we can
always resort to their evolutionary representative in order to be
within this framework. It should be noted, however, that when $X =
Q^a \pa_a$ is the evolutionary representative of a vector field
$X_0$ with a horizontal component, say $X_0 = \xi^i \pa_i + \phi^a
\pa_a$ (we will assume at least one of the $\xi^i$ is not
identically zero), we are not at all guaranteed that the vertical
vector field $\wt{X}$ whose standard prolongation is
gauge-equivalent to $Y = X^{(n)}_\la$ is the evolutionary
representative of some Lie-point  vector field $\wt{X}_0$ in $M$.
This point is usually overlooked in the literature about
$\mu$-prolongations and $\mu$-symmetries, and we want to discuss
it in some detail (it will suffice to consider first
prolongations, i.e. $n=1$).

For $X_0$ as above, we will have $Q^a = \phi^a - u^a_i \xi^i$; thus
$$ Y \ = \ Q^a \, \pa_a \ + \ [(D_i Q^a) + (\La_i)^a_{\ b} Q^b ] \, \pa_a^i \ . $$
We will look for $\wt{X}_0$ in the form
\beq\label{eq:wtx0} \wt{X}_0 \ = \ \eta^i \, \pa_i \ + \ \vth^a \, \pa_a \ ; \eeq
its evolutionary representative is
$$ \wt{X} \ = \ P^a \, \pa_a \ = \ (\vth^a - u^a_j \eta^j ) \, \pa_a \ . $$
The standard prolongation of this is
\begin{eqnarray*} Z &=& P^a \, \pa_a \ + \ (D_i P^a) \, \pa_a^i \\
&=&  (\vth^a - u^a_i \eta^i ) \ + \ [ (D_i \vth^a) - u^a_{j,i} \eta^j - u^a_j (D_i \eta^j)] \pa_a^i \ . \end{eqnarray*}

We note that dependence on $u^a_i$ is explicit in both $P^a$ and
$Q^a$, as $\xi^i , \eta^i, \phi^a,\vth^a$ only dpeend on $x$ and
$u$; thus requiring that $\wt{X}$ be the gauge transform via a map
$A$ of $X$, i.e. that $Q^a = A^a_{\ b} P^b$ amounts to the
requirements
\begin{eqnarray}
\phi^a &=& A^a_{\ b} \ \vth^b \ , \label{eq:muinv1} \\
u^a_j \, \xi^j &=& A^a_{\ b} \, u^b_i \ \eta^i \ .
\label{eq:muinv2} \end{eqnarray} Note that we should also require
the same holds for $Y$ and $Z$; this would fix the relation
between $A$ and the $\La_i$ as discussed above. We will thus
suppose $A$ is known.

Now, it is clear that if we know $A$ then \eqref{eq:muinv1}
immediately provides \beq \label{eq:vth} \vth^a \ = \
(A^{-1})^a_{\ b} \ \phi^b \ . \eeq On the other hand,
\eqref{eq:muinv2} does not provide $\eta^i$ with equal ease. First
of all, we note that albeit we have sums on the indices $i$ (in
the right hand side) and $j$ (in the left hand side), these
indices are not acted upon by the map $A$, so that the equality
must hold for any index, and we can write
$$ u^a_i \, \xi^i \ = \ \sum_b \ A^a_{\ b} \, u^b_i \ \eta^i $$
with no sum on $i$. This is a system of equations (indexed by
$a$), which should hold for any value of the $u^a_i$; thus (unless
the $\xi^i$ are all zero) it may be satisfied only if $A$ is
diagonal. But if $A$ is diagonal, then $\La_i = (D_i A) A^{-1}$ is
also diagonal. We conclude that a necessary condition for $\wt{X}$
to be the evolutionary representative of a vector field $\wt{X}_0$
is that the $\La_i$ (and $A$) are diagonal.

If this is the case, say with $A = \mathtt{diag} (\a_1 , ... ,
\a_p )$, then \eqref{eq:muinv2} splits into $p$ scalar equations
(for $b=1,...,p$),
$$ \xi^i \ = \ \a_b \ \eta^i \ ; $$
it suffices to recall we have assumed $A$ is invertible, and hence
all the $\a_k$ are nonzero, to conclude that in this case
\beq \label{eq:eta} \eta^i \ = \ (\a_b )^{-1} \ \xi^i \ .
\eeq
Note that unless $\a_1=...=\a_p$, the equations for different $b$
would give different determinations of the $\eta^i$; thus $\eta^i$
(and hence the vector field $\wt{X}_0$) can be determined {\it only}
for $A$ a multiple of the identity, $A = \la I$.

Our discussion can also be conducted in a slightly different way
(reaching of course the same conclusions). That is, in order to
reconstruct the original Lie-point vector field $X$ from its
evolutionary representative $X_v$ it suffices, in view of the
explicit expression for $Q^a = \phi^a - u^a_i \xi^i$ and of the
fact $\phi$ and $\xi$ do not depend on $u$ derivatives, to write
(no sum on $a$)
\beq\label{eq:reconst} \xi^i \ = \ - \ \frac{\pa Q^a}{\pa u^a_i} \
; \ \ \phi^a \ = \ Q^a \ + \ u^a_i \, \frac{\pa Q^a}{\pa u^a_i} \
. \eeq Note that we do {\it not} sum on $a$ in the formula for
$\xi^i$; we have instead that the result is the same for any $a$.

When considering $\wt{X} = P^a \pa_a$ with $P = A^{-1} Q$, as seen
above, if we want to write $P$ as $P^a = \vth^a - u^a_i \eta^i$ we
run in troubles in the determination of $\eta^i$. In fact, we
immediately have
$$ P^a \ = \ (A^{-1})^a_{\ b} \, \phi^b \ - \ [(A^{-1})^a_{\ b} \,
u^b_i] \, \xi^i \ . $$ According to the prescription
\eqref{eq:reconst}, we immediately obtain $$ \vth^a \ = \
(A^{-1})^a_{\ b} \, \phi^b \ , $$ as it should be. But for what
concerns $\eta$, the same \eqref{eq:reconst} would provide (again,
no sum on $a$) $$ \eta^i \ = \ \frac{\pa P^a}{\pa u^a_i} \ = \
(A^{-1})^a_{\ a} \, \xi^i \ . $$ This would provide different
results for different choices of $a$; unless of course all the
diagonal element of $A^{-1}$ are the same.

Summarizing, if we have vector fields with
horizontal components, then the diagram \eqref{diag:mu} applies to
the evolutionary representative, and $\wt{X}$ will in general be a
generalized vector field which is not the evolutionary
representative of a Lie-point vector field $\wt{X}_0$ in $M$; and
this unless $A$ is diagonal (which in turn requires the $\La_i$
are all diagonal). In this case, $\wt{X}_0$ is identified by
\eqref{eq:wtx0} with \eqref{eq:vth} and \eqref{eq:eta}.

\section{Collective and combined twisted symmetries}

\subsection{Collective twisted symmetries: $\s$-symmetries}
\label{sec:sigsymm}

So far we have considered vector fields with modified prolongation
rule, but each vector field was however prolonged independently.
It is possible to consider a different modification of the
prolongation operation, which makes sense on sets of vector
fields; these must be in involution (in Frobenius sense).

It should be noted that so far these have been studied (and shown
to be useful) only in the context of ODEs; we will thus restrict
to this setting, i.e. have only one independent variable, denoted
with $x$.

We will thus consider a set of vector fields $X_\a$ ($\a = 1,...,
r$) on $M$, written in coordinates as \beq X_{\a} \ = \ \xi_{\a}
\, \frac{\pa}{\pa x} \ + \ \phi_{\a}^a \, \frac{\pa}{\pa u^a} \ ;
\eeq and satisfying the involution relations \beq \[ X_{\a} \, ,
\, X_{\b} \] \ = \ f_{\a \b}^\ga \ X_{\ga} \ , \eeq where the
$f_{\a \b}^\ga : M \to \R$ are smooth functions. Thus we deal with
a {\it Lie module} of vector fields in $M$.

We will write the $\s$-prolongations of these -- to be
characterized by the following formula \eqref{eq:sigprol} -- as
\beq Y_{\a} \ = \ \xi_{\a} \, \frac{\pa}{\pa x} \ + \
\psi_{\a,(k)}^a \, \frac{\pa}{\pa u^a_{(k)}} \ ; \eeq note that
the index $k$ refers to the order of derivations.

Here $\psi_{\a,0}^a = \phi_\a^a$, and the $\psi_{\a,(k)}^\a$
satisfy the {\it $\s$-prolongation} formula \beq
\label{eq:sigprol} \psi_{\a,(k+1)}^a \ = \ \( D_x \,
\psi_{\a,(k)}^a \ - \ u^a_{(k+1)} \ D_x \xi_\a \) \ + \ \s_\a^{\
\b} \ \( \psi_{\b,(k)}^a \ - \ u^a_{(k+1)} \xi_\b \) \ ; \eeq here
the $\s_\a^{\ \b}$ are smooth real functions on $J^1 M$.

We stress that while $\mu$-prolongations were mixing different
components of the same vector field, the $\s$-prolongations mix
the same component (vector index $a$) of different vector fields.
In particular, this implies that $\s$-prolongations make sense for
sets (actually, as noted above, Lie modules) of vector
fields.\footnote{In the case of a trivial module, generated by a
single vector field, we are reduced to $\la$-prolongations.}

\medskip\noindent
{\bf Lemma 8.} {\it The $\s$-prolonged vector fields satisfy \beq
\label{eq:commsig} \[ Y_\a , D_x \] \ = \ \s_\a^{\ \b} \, Y_\b \ -
\ \( D_x \xi_\a \, + \, \s_\a^{\ \b} \xi_\b \) \, D_x \ . \eeq
Conversely, if the vector fields $Y_\a$ satisfy
\eqref{eq:commsig}, they are $\s$-prolonged.}

\medskip\noindent
{\bf Proof.} This follows by a simple computation, making use of
\eqref{eq:sigprol}. In fact, with the notation introduced earlier
on, we have
\begin{eqnarray*}
[Y_\a , D_x ] &=& \( \psi^a_{\a,(k+1)} \ - \ D_x \psi^a_{\a , (k)}
\) \, \pa_a^k \ - \ (D_x \xi_\a) \, \pa_x \\
&=& \( - u^a_{(k+1)} \ (D_x \xi_\a ) \ + \ \s_\a^{\ \b} \ (
\psi^a_{\b,(k)} \, - \, u^a_{(k+1)} ) \) \, \pa_a^k \ - \ (D_x
\xi_\a ) \, \pa_x \\
&=& - \, (D_x \xi_\a) \, D_x \ + \ \s_\a^{\ \b} \, (Y_\b - \xi_\b
\pa_x) \ - \ \s_\a^{\ \b} \xi_\b \, (D_x - \pa_x ) \\
&=& - \, \( (D_x \xi_\a) \, + \, \s_\a^{\ \b} \, \xi_\b \) \, D_x
\ + \ \s_\a^{\ \b} \, Y_\b \ . \end{eqnarray*} The converse
statement follows performing the computation in reverse order.
\EOP

\medskip\noindent
{\bf Remark 20.} It should be stressed that for a generic choice
of $\s_\a^{\ \b}$, the $\s$-prolonged vector fields $Y_\a$ would
not satisfy the same involution relations as the $X_\a$. It can be
shown (see Theorem 2 in \cite{Sprol}; the proof is based on a
relatively long computation) that this is the case if and only if
$\s$ satisfies \beq \[ \( Y_\a (\s_\b^{\ \ga} ) - Y_\b (\s_\a^{\
\ga}) \) \, + \, \( (D_x f_{\a \b}^\ga ) + \s_\a^{\ \eta}  f_{\eta
\b}^\ga  - \s_\b^{\ \eta}  f_{\eta \a}^\ga  -  f_{\a \b}^\eta
\s_\eta^{\ \ga} \) \] \phi^a_\ga \, = \, 0 \, ; \eeq a sufficient
(but not necessary) for this to hold is to have \beq \(
 Y_\a (\s_\b^{\ \ga} ) - Y_\b (\s_\a^{\ \ga}) \) \, + \, \( (D_x
f_{\a \b}^\ga ) +  \s_\a^{\ \eta}  f_{\eta \b}^\ga  - \s_\b^{\
\eta}  f_{\eta \a}^\ga  -  f_{\a \b}^\eta  \s_\eta^{\ \ga} \) \, =
\, 0 \, . \eeq In the case the $X_\a$ commute, so that $f_{\a
\b}^\ga = 0$, this further reduces to \beq
 Y_\a (\s_\b^{\ \ga} ) - Y_\b (\s_\a^{\ \ga}) \ = \ 0 \ . \eeq

Note also that $\s$-prolongation can be defined irrespective of
the involution properties of the $Y_\a$; however, reduction via
$\s$-symmetries (see below) requires that the $Y_\a$ are an
involution system\footnote{Not necessarily with the same
involution properties; in particular, we could add new vector
fields in $J^n M$ to close the system under commutation and obtain
an involution system.}, and this is not guaranteed {\it a priori},
unless the above conditions are satisfied. \EOR
\bigskip

A set $\{ X_\a , \a = 1,...,r \}$ of vector fields in involution
will be said to be a $\s$-symmetry module (or algebra if the
$f_{\a \b}^\ga$ are constant) of the system $\De$ if their
$\s$-prolongations $Y_\a$ satisfy \beq Y_\a \ : \ S_\De \to \T
S_\De \ . \eeq

As we have seen above, the possibility of using symmetries for
reduction of ODEs is ultimately based on the IBDP; the same will
hold true (as discussed below) for $\s$-symmetries, so that we
need to ascertain if $\s$-prolonged (sets of) vector fields enjoy
the IBDP. This is actually the case.

\medskip\noindent
{\bf Lemma 9.} {\it A set $\{ Y_1 , ... , Y_r \}$ of
$\s$-prolonged vector fields in involution enjoy the IBDP
property. That is, if $\eta$ and $\zeta_{(k)}$ are independent
common differential invariants for the $Y_\a$, so is
$$ \zeta_{(k+1)} \ := \ \frac{D_x \zeta_{(k)} }{D_x \eta} \ . $$}

\medskip\noindent
{\bf Proof.} We note that
$$ Y_\a (\zeta_{(k+1)} \ = \ \frac{[Y_\a (D_x \zeta_{(k)}] \cdot
(D_x \eta) \ - \ (D_x \zeta_{(k)} ) \cdot [Y_\a (D_x \eta)]}{(D_x
\eta)^2} \ := \ \frac{\Theta}{(D_x \eta)^2} \ ; $$ so we just have
to show that $\Theta = 0$.

This follows by a straightforward computation (going along the
lines of the one met in the proofs of Lemma 1, Lemma 3 and Lemma
5), in which we make use of the assumption that $\eta$ and
$\zeta_{(k)}$ are common (scalar) differential invariants for the
$Y_\a$, so that $Y_\a \eta) = 0 = Y_\a (\zeta_{(k)}$, and of
\eqref{eq:commsig}. In fact, we have
\begin{eqnarray*}
\Theta &=& [Y_\a (D_x \zeta_{(k)}] \cdot (D_x \eta) \ - \ (D_x
\zeta_{(k)} ) \cdot [Y_\a (D_x \eta)] \\
&=& \( D_x \(Y_\a (\zeta_{(k)}) \) \ + \  [Y_\a , D_x]
(\zeta_{(k)}) \) \cdot (D_x \eta) \\ & & \ \ \ - \ (D_x
\zeta_{(k)} ) \cdot \( D_x \( Y_\a
(\eta) \) \ + \ [Y_\a, D_x] (\eta) \) \\
&=& \( [Y_\a , D_x] (\zeta_{(k)}) \) \cdot (D_x \eta) \ - \ (D_x
\zeta_{(k)} ) \cdot \( [Y_\a, D_x] (\eta) \) \\
&=& - \, (D_x \xi_\a + \s_\a^{\ \b} \xi_\b) \cdot (D_x \zeta_{(k)}
) \cdot (D_x \eta) \\ & & \ \ \  + \ (D_x \zeta_{(k)} ) \cdot (D_x
\xi_\a +
\s_\a^{\ \b} \xi_\b) \cdot (D_x \eta) \\
&=& 0 \ . \end{eqnarray*} This concludes the proof. \EOP

\medskip\noindent
{\bf Remark 21.} Note that if $\eta$ and $\zeta_{(k)}$ are of
order $k$, then $\zeta_{(k+1)}$ is of order $k+1$, and hence
independent of $\eta$ and $\zeta_{(k)}$. Thus if we have a
complete basis for the set of differential invariants of order
zero and one for the $Y_\a$, we can generate a basis of invariants
of all orders just by differentiation. Note also that we have here
{\it assumed} the $Y_\a$ are in involution; as mentioned in Remark
20, this is in general not granted. \EOR
\bigskip

The fact that differential invariants can be generated by means of
differentiation allows to implement the same reduction strategy as
for standard (or, for that matter, $\lambda$) symmetries. That is,
we pass to symmetry-adapted variables, in which the vector fields
have a trivial expression, and the differential equations system
is simply independent of the new dependent variables. More
precisely, in the simplest setting we have the following

\medskip\noindent
{\bf Lemma 10.} {\it Let $\mathcal{X} = \{ X_1,...,X_r \}$ be a
system (of rank $r$) of vector fields in involution on $M = \R
\times \R^p$; let $\mathcal{Y} = \{ Y_1 , ... , Y_r \}$ be their
$\s$-prolongation on $J^n M$, still in involution and of rank $r$;
and let the system $\De$ of $p$ ordinary differential equations of
order $n > 1$ admit $\mathcal{X}$ as a $\s$-symmetry set. Assume
moreover $r=p$. Then $\De$ can be reduced to a system of $p$
differential equations of order $n-1$.}

\medskip\noindent
{\bf Proof.} The $\mathcal{\X}$ span a distribution of rank $r=p$
in $M$, which is of dimension $p+1$; hence they admit one common
invariant. By assumption, the distribution generated by
$\mathcal{Y}$ has also dimension $r=p$; thus at each order
$k=1,...,n$ there exist exactly $p$ independent differential
invariants of order $k$.

We pass to symmetry adapted coordinates $(y; w^1 , ... , w^p)$
such that $X_\a (y) = 0$ for all $\a = 1,...,r$. Thus the
geometrical invariant is simply $\eta (y,w) = y$. Call $\zeta_a
(y,w,w_{(1)})$ (here $a=1,...,p$) the common differential
invariants of order one, i.e. the functions $\zeta^a : J^1 M \to
\R$ satisfying $Y_\a (\zeta^a ) = 0$ for all $\a$ and $a$.

Thanks to Lemma 9, the common differential invariants of order
$k+1$ are obtained as $$ \zeta^a_{(k)} \ := \ D_x^k \ \zeta^a \ .
$$ With a new change of dependent variables, we choose these as $$
z^a \ = \ \zeta^a (y;w,w_{(1)} ) \ . $$ The system of differential
equations will be written in these new variables as
$$ F^a (y; z , ... , z_{(n-1)} ) \ = \ 0 \ ; $$ this realizes the
reduction to a system of equations of order $n-1$. \EOP

\medskip\noindent
{\bf Remark 22.} A similar result holds for $r < p$ (and also for
equations not all of the same order); in this case, only $r$ of
the $p$ differential equations are reduced to a lower order; see
Theorem 4 in \cite{Sprol}. \EOR

\medskip\noindent
{\bf Remark 23.} We also mention that the case of dynamical
systems, i.e. systems of first order (autonomous) ODEs presents
some special difficulties as far as $\s$-symmetries as concerned;
these are due to the fact such a simple framework does actually
introduce several degeneracies with respect to the general
discussion. For a discussion of $\s$-symmetries (and their
application) in the context of dynamical systems, see
\cite{SprolDS}. \EOR

\medskip\noindent
{\bf Remark 24.} It should be stressed that, at difference with
the standard symmetry case, obtaining solutions to the original
equations from solutions to the reduced one (the so called {\it
reconstruction problem}) is not trivial in this case.

In fact, we do not only have to invert the change of coordinates
$(x,u) \to (y,w)$, but also have to solve the equation
$$ z^a (y) \ = \ \zeta^a [y ; w (y) , w_{(1)} (y) ] \ , $$
which now has to be considered as a system of $r$ first order
differential equations for $w^a (y)$.

In other words, while for standard symmetries the reconstruction
problem reduces to quadratures, in the present case one is faced
with an auxiliary system of (in general, nonlinear and non
autonomous) differential equations. Needless to say, there is no
guarantee we will be able to solve this system. \EOR
\bigskip

In the case of $\la$-symmetries, we have seen that there is a
simple geometric relation between $\la$-prolonged and standardly
prolonged vector fields, see Lemma 6. The same holds here, but as
in this case $\s$-prolongations acts on sets of vector fields, the
comparison should be made between $\s$ and standard prolongations
of sets of vector fields.

\medskip\noindent
{\bf Lemma 11.} {\it Let $\mathcal{X} = \{ X_1 , ... , X_r \}$ be
a set of vector fields on $M$; and let the vector fields
$\mathcal{Y} = \{ Y_1 , ... , Y_r \}$ on $J^n M$ be their
$\s$-prolongation. Consider also the set $\mathcal{W} = \{ W_1 ,
... , W_r \}$ of vector fields on $M$ given by $W_\a = A_\a^{\ \b}
X_\b$, with $A$ a nowhere singular matrix function on $M$; and let
the vector fields $\mathcal{Z} = \{ Z_1 , ... , Z_r \}$ on $J^n M$
be their standard prolongation. Then, provided $A$ and $\s$ are
related by \beq\label{eq:L11} A^{-1} \ D_x A \ = \ \s \ , \eeq we
also have $Z_\a = A_\a^{\ \b} Y_\b$.}

\medskip\noindent
{\bf Proof.} It will suffice to consider first prolongations. In
coordinates and with the usual shorthand notation,
\begin{eqnarray*} X_\a &=& \xi_\a \, \pa_x \ + \ \phi^a_\a \,
\pa_a \ , \\ Y_\a &=& X_\a \ + \ \( (D_x \phi^a_\a \, - \, u^a_x
\, D_x \xi_\a) \ + \ \s_\a^{\ \b} \, (\phi^a_\b \, - \, u^a_x \,
\xi_\b ) \) \, \pa_a^1 \ ; \\
W_\a &=& \chi_\a \, \pa_x \ + \ \eta^a_\a \, \pa_a \ , \\ Z_\a &=&
W_\a \ + \ (D_x \eta^a_\a \, - \, u^a_x \, D_x \chi_\a) \, \pa_a^1
\ . \end{eqnarray*}

If now we require $W = A X$, i.e.
$$ \chi_\a \ = \ A_\a^{\ \b} \, \xi_\b \ ; \ \
\eta^a_\a \ = \ A_\a^{\ \b} \, \phi^a_\b \ , $$ we immediately get
\begin{eqnarray*}
D_x \chi_\a &=& (D_x A_\a^{\ \b}) \, \xi_\b \ + \ A_\a^{\ \b} \,
(D_x \xi_\b) \ , \\
D_x \eta^a_\a &=& (D_x A_\a^{\ \b}) \, \phi^a_\b \ + \ A_\a^{\ \b}
\, (D_x \phi^a_\b) \ . \end{eqnarray*} Inserting these in the
expression for $Z$, we get \begin{eqnarray*} Z &=& A_\a^{\ \b} \
\[ \xi_\b \, \pa_x \ + \ \phi^a_\b \, \pa_a \right. \\
& & \left. \ \ \ + \ \( (D_x \phi^a_\b - u^a_x D_x \xi_\b) \ + \
(A^{\-1} \, D_x A)_\a^{\ \b}
\, (\phi^a_\b - u^a_x \xi_\b ) \) \] \, \pa_a^1 \\
&=& A_\a^{\ \b} \, Y_\b \ + \ \[ (D_x A)_\a^{\ \b} \ - \ (A
\s)_\a^{\ \b} \] \, (\phi^a_\b - u^a_x \xi_\b ) \, \pa_a^1 \ .
\end{eqnarray*} Thus, provided $A$ and $\s$ satisfy $D_x A = A
\s$, and hence (recalling $A$ is nowhere singular) satisfy
\eqref{eq:L11}, we have that $W = A X$ leads to $Y = A Z$. \EOP

\medskip\noindent
{\bf Corollary.} {\it With the setting and notation of Lemma 11,
denote by $\mathcal{D}$ the distribution generated by
$\mathcal{Y}$, and denote by $\^\mathcal{D}$ the distribution
generated by $\mathcal{Z}$. Then the distributions $\^\mathcal{D}$
and $\mathcal{D}$ coincide.}

\medskip\noindent
{\bf Proof.} This follows immediately from Lemma 11. \EOP

\medskip\noindent
{\bf Remark 25.} The relations between the vector fields $X_i$ and
$W_i$, and their (respectively, $\s$ and standard) prolongations,
as given by Lemma 11, can be summarized in the form of a
commutative diagram:
$$ \matrix{
\{ X_i \} & \mapright{A} & \{ W_i \} \cr \mapdown{\s-prol} & &
\mapdown{prol} \cr \{ Y_i \} & \mapright{A} & \{ Z_i \} \cr} $$
The relation between $A$ and $\s$ is given by \eqref{eq:L11}. \EOR

\medskip\noindent
{\bf Remark 26.} Lemma 11 stipulates that the relation between $A$
and $\s$ is given by \eqref{eq:L11}; we note that if we look at this
as an equation for $A$ with a given $\s$, the solution is in
general not unique (an explicit example is provided in
\cite{Sprol}). Moreover -- as in Remarks 10 and 16 -- unless $\s = D_x S$ for some local matrix function $S$ defined on $M$, $A$ can and will be a nonlocal
function,
$$ A \ = \ \exp \[ \int \s \ d x \] \ . $$
Finally, we note that the sets $\{Y_i \}$ and $\{ Z_i \}$
considered in Lemma 11 will in general have different involution
properties. \EOR

\subsection{Combining simple and collective twisted symmetries: $\chi$-symmetries}
\label{sec:chisymm}

We have seen that $\mu$-prolongations (and hence symmetries) are
associated to matrices $\La_i$ -- or to a single matrix $\La$ when
we consider ODEs -- acting on the vector indices ($a,b,...$) of
the coefficients of each vector field; on the other hand,
$\s$-prolongations (and hence symmetries) are associated to a
matrix $\s$ acting on the Lie module indices ($\a,\b,...$)
denoting the different vector fields.

It has to be expected that the two different ``twisting'' of the
prolongation operation can be combined with no problem, as they
act on different structures. This is indeed the case, and the
``combined'' twisted prolongation operation (and hence symmetries)
has been denoted as {\it $\chi$-prolongation} \cite{Sprol}. Note
that $\s$-prolongations are (so far) only defined in the framework
of a single independent variable, i.e. ODEs; hence the same
applies for $\chi$-prolongations.

In order to discuss such combined twisted prolongations, we will
slightly change our notation; moreover, we will limit to discuss
vertical vector fields.\footnote{These can be thought as the evolutionary
representative of general vector fields, see above; hence the coefficients
$\phi$, $\^\phi$, $\wt{\phi}$ are not only functions of the $x$ and $u$ but
could depend -- linearly -- on the $u^a_i$ as well.}

We will thus consider a set of vector fields $W_\a$ on $M$,
written as
$$ W_\a \ = \ \phi^a_\a \, \pa_a \ ; $$
their standard prolongations will be denoted as $Z_\a$ and will be
written as
$$ Z_\a \ = \ \psi^a_{\a,(k)} \, \pa_a^k \ , $$
with $\psi^a_{\a,(0)} = \phi^a_\a$ and the $\psi^a_{\a,(k)}$
obeying the standard prolongation formula \eqref{eq:prolformODEs}.

We will then consider another set of vector fields $\^X_\a$ on
$M$, written as
$$ \^X_\a \ = \ \^\phi^a_\a \, \pa_a \ ; $$
their $\mu$-prolongations will be denoted as $\^Y_\a$ and will be
written as
$$ \^Y_\a \ = \ \^\psi^a_{\a,(k)} \, \pa_a^k \ , $$
with $\^\psi^a_{\a,(0)} = \^\phi^a_\a$ and the $\^\psi^a_{\a,(k)}$
obeying the $\mu$-prolongation formula \eqref{eq:muprol} (with a
single independent variable, and hence a single $\La$).

Finally we also consider a third set of vector fields $\wt{X}_\a$
on $M$, written as
$$ \wt{X}_\a \ = \ \wt{\phi}^a_\a \, \pa_a \ ; $$
their $\s$-prolongations will be denoted as $\wt{Y}_\a$ and will
be written as
$$ \wt{Y}_\a \ = \ \wt{\psi}^a_{\a,(k)} \, \pa_a^k \ , $$
with $\wt{\psi}^a_{\a,(0)} = \wt{\phi}^a_\a$ and the
$\wt{\psi}^a_{\a,(k)}$ obeying the $\s$-prolongation formula
\eqref{eq:sigprol}.

The three sets are related by \beq \^X_\a \ = \ A \, W_\a \ ; \ \
\wt{X}_\a \ = \ B_\a^{\ \b} \, W_\b \ . \eeq These should be meant
as \beq \^\phi^a_\a \ = \ A^a_{\ b} \, \phi^b_\a \ , \ \
\wt{\phi}^a_\a \ = \ B_\a^{\ \b} \, \phi^a_\b \ . \eeq

The relations between these different sets of vector fields --
which were discussed in Sections \ref{sec:musymm} e
\ref{sec:sigsymm} -- are then summarized in the following diagram
\beq \label{diag:chi1} \matrix{ \{ \^X_\a \} & \mapleft{A} & \{
W_\a \} & \mapright{B} & \{ \wt{X}_\a \} \cr
 & & & & \cr
\mapdown{\mu-prol} & & \mapdown{prol} & & \mapdown{\s-prol} \cr
 & & & & \cr
 \{ \^Y_\a \} & \mapleft{A} & \{ Z_\a \} & \mapright{B} & \{ \wt{Y}_\a \} \cr} \eeq

The invertible matrices $A$ and $B$ are related to the matrices
$\s$ and $\La$ (the latter identifying $\mu$ via $\mu  = \La \d
x$) by
\begin{eqnarray}
\La &=& A \ (D_x A^{-1} ) \ = \ - \ (D_x A) \ A^{-1} \ , \label{eq:recA} \\
\s  &=& B \ (D_x B^{-1} ) \ = \ - \ (D_x B) \ B^{-1} \ .
\label{eq:recB} \end{eqnarray}
The usual remarks about solution of these equations for given $\s$ and $\La_i$, and the locality of the matrix functions $A$ and $B$ thus determined, would apply; see Remarks 10, 16 and 26.

We can now apply the $B$ map on the set of the $\^X_\a$, or the
$A$ map on the set of the $\wt{X}_\a$, and obtain a new set of
vector fields \beq P_\a \ = \ A \ \wt{X}_\a \ = \ B_\a^{\ \b} \
\^X_\b \ . \eeq Note that the $P_\a$ will be written in
coordinates as \beq P_\a \ = \ \Phi^a_\a \ \pa_a \ = \ A^a_{\ b} \
B_\a^{\ \b} \ \phi^b_\b \ \pa_a \ . \eeq It is obvious that the
order in which the $A$ and $B$ maps are applied is inessential, as
they act on different sets of indices.

We can now apply the same $A$ and $B$ on the (respectively, $\mu$
and $\s$) prolongations of the $\^X_\a$ and $\wt{X}_\a$, obtaining
new vector fields \beq Q_\a \ = \ A \ \wt{Y}_\a \ = \ B_\a^{\ \b}
\ \^Y_\b \eeq in $J^n M$; these are written in coordinates as \beq
\label{eq:Q} Q_\a \ = \ \Psi^a_{\a,(k)} \ \pa_a^k \ = \ A^a_{\ b}
\ B_\a^{\ \b} \ \psi^b_{\b,(k)} \ \pa_a^k \ , \eeq where of course
$\Psi^a_{\a,(0)} = \Phi^a_\a$.

We will say that the $\{ Q_\a \}$ are the $\chi$-prolongations of
the $P_\a$. In order to keep track of the different components of
this combined prolongation, we will also write $\chi = (\mu, \s)$.
Note that $\chi$-prolongations (just like $\s$-prolongations) are
defined on sets of vector fields, not on single ones.

We can characterize $\chi$-prolongations in terms of the recursion
relations satisfied by the coefficients $\Psi^a_{\a,(k)}$
appearing in \eqref{eq:Q}. These recursion relations are easily
computed in terms of those for the $\psi^a_{\a,(k)}$ and the
matrices $A$ and $B$.

\medskip\noindent
{\bf Lemma 12.} {\it The coefficients $\Psi^a_{\a,(k)}$ in the
$\chi$-prolonged vector field $Q$ \eqref{eq:Q}, with $A$ and $B$
obeying \eqref{eq:recA} and \eqref{eq:recB}, satisfy the
$\chi$-prolongation formula \beq\label{eq:chiprol}
\Psi^a_{\a,(k+1)} \ = \ D_x \, \Psi^a_{\a,(k)} \ + \ \La^a_{\ b}
\, \Psi^b_{\a,(k)} \ + \ \s_\a^{\ \b} \, \Psi^a_{\b,(k)} \ . \eeq}

\medskip\noindent
{\bf Proof.} This just follows by a simple computation. We know
that for any $k \ge 0$, $\Psi^a_{\a,(k)}$ satisfies
$$ \Psi^a_{\a,(k)} \ = \ A^a_{\ b} \, B_\a^{\ \b} \, \psi^b_{\b,(k)} \ ; $$
we write this, inverting the relation, as
$$ \psi^a_{\a,(k)} \ = \ (A^{-1})^a_{\ b} \, (B^{-1})_\a^{\ \b} \ \Psi^b_{\b,(k)} \ . $$
In turn, we have
$$ \psi^a_{\a,(k+1)} \ = \ D_x \psi^a_{\a,(k)} \ . $$
It follows from these, with a shorthand notation which omits to
explicitly write indices, and recalling that matrices of type $A$
and $B$ commute and act on different sets of indices, that
\begin{eqnarray*}
\Psi_{(k+1)} &=& A \, B \, \psi_{(k+1)} \\
&=& A \, B \, (D_x \psi_{(k)} ) \\
&=& A \, B \, \( D_x ( A^{-1} \, B^{-1} \ \Psi_{(k)} ) \) \\
&=& A  B \, \[ ( D_x A^{-1})  B^{-1}  \Psi_{(k)}  +  A^{-1}  (
D_x B^{-1})  \Psi_{(k)} \, + \,
A^{-1} B^{-1} (D_x \Psi_{(k)} ) \] \\
&=& A  B \, \[ ( D_x A^{-1})  B^{-1}  \Psi_{(k)}  +  A^{-1} (
D_x B^{-1} ) \Psi_{(k)} \, + \,
A^{-1}  B^{-1}  (D_x \Psi_{(k)} ) \] \\
&=& A \, ( D_x A^{-1}) \, \Psi_{(k)} \ + \ B \, ( D_x B^{-1} ) \,
\Psi_{(k)} \ + \ (D_x \Psi_{(k)} ) \ . \end{eqnarray*} Recalling
now \eqref{eq:recA} and \eqref{eq:recB}, this is rewritten as
$$ \Psi_{(k+1)} \ = \ (D_x \Psi_{(k)} ) \ + \ \La \, \Psi_{(k)} \ + \ \s \, \Psi_{(k)} \ ; $$
reintroducing indices, we have precisely \eqref{eq:chiprol}. \EOP
\bigskip

With the introduction of $\chi$-prolongations, the diagram
\eqref{diag:chi1} is then complemented by the follow diagram: \beq
\label{diag:chi2} \matrix{ \{ \^X_\a \} & \mapright{B} & \{ P_\a
\} & \mapleft{A} & \{ \wt{X}_\a \} \cr
 & & & & \cr
\mapdown{\mu-prol} & & \mapdown{\chi-prol} & & \mapdown{\s-prol}
\cr
 & & & & \cr
 \{ \^Y_\a \} & \mapright{B} & \{ Q_\a \} & \mapleft{A} & \{ \wt{Y}_\a \} \cr} \eeq


We will as usual say that the equations $\De$ of order $n$ admits
$P$ as a $\chi$-symmetry if its $\chi$-prolongation of order $n$,
$Q$, satisfies $Q : S_\De \to \T S_\De$.

It remains to be discussed how ODEs are reduced under
$\chi$-symmetries. This depends on $\chi$-prolonged vector fields
enjoying the IBDP or otherwise. In fact, we know that
$\La$-prolonged and $\s$-prolonged vector fields enjoy the IBDP
property, and it is easy to conclude that also $\chi$-prolonged
vector fields -- provided the matrix $\La$ is a multiple of the
identity -- enjoy the IBDP property. We will actually, for the
sake of completeness, prove this by explicit computation.

\medskip\noindent
{\bf Lemma 13.} {\it Let the vector fields $Q_\a$ in $J^n M$ be
the $\chi$-prolongation of vector fields $P_\a$ in $M$, with $\chi
= (\mu, \s)$ and $\mu  = \La \d x$. Assume moreover that $\La =
\la I$, with $\la : J^1 M \to \R$. Then the set $\mathcal{Q} = \{
Q_\a \}$ has the IBDP.}

\medskip\noindent
{\bf Proof.} We proceed as, and with the same notation as, in the
proofs of Lemmas 1, 3, 5 and 9. Using
$$ Q_\a \ = \ \Psi^a_{\a,(k)} \, \pa_a^k \ ; \ \ \ D_x = \pa_x \ + \ u^a_{(k+1)} \, \pa_a^k \ , $$ we immediately get
$$ [Q_\a , D_x ] \ = \ \( \Psi^a_{\a,(k+1)} \ - \ (D_x \Psi^a_{\a,(k)}) \) \ \pa_a^k \ . $$
With \eqref{eq:chiprol}, this yields
$$ [Q_\a , D_x ] \ = \ \( \La^a_{\ b} \, \Psi^b_{\a,(k)} \ + \ \s_\a^{\ \b} \, \Psi^a_{\b,(k)} \) \ \pa_a^k \ . $$
Using now the hypothesis $\La = \la I$, this gives simply
$$ [Q_\a , D_x ] \ = \ \( \la \, \Psi^a_{\a,(k)} \ + \ \s_\a^{\ \b} \, \Psi^a_{\b,(k)} \) \ \pa_a^k \ = \
\la \, Q_\a \ + \ \s_\a^{\ \b} \, Q_\b \ . $$ We will introduce
the matrix $$ \rho \ := \ \la \, I \ + \ \s \ ; $$ the previous
relation is thus written as \beq \label{eq:chisomm} [Q_\a , D_x ]
\ = \  \rho_\a^{\ \b} \ Q_\b \ . \eeq

To show the IBDP, we have to show that for any common differential
invariants $\eta$ and $\zeta_{(k)}$, and for an $\a$, it results
$Q_\a [ (D_x \zeta_{(k)} ) / (D_x \eta) ] = 0$, i.e. that
$$ \Theta_\a \ := \ \( Q_\a (D_x \zeta_{(k)} ) \) \cdot (D_x \eta)
\ - \ (D_x \zeta_{(k)} ) \cdot [Q_\a (D_x \eta)] \ = \ 0 \ . $$
This can be rewritten as
\begin{eqnarray*}
\Theta_\a &=& \( Q_\a (D_x \zeta_{(k)} ) \) \cdot (D_x \eta) \ - \
(D_x \zeta_{(k)} ) \cdot [Q_\a (D_x \eta)] \\
&=& \( [Q_\a,D_x] (\zeta_{(k)} ) \, - \, D_x ( Q_\a (\zeta_{(k)} ) ) \)
\cdot (D_x \eta) \\
& & \ \ \ \ \ - \
(D_x \zeta_{(k)} ) \cdot \( [Q_\a,D_x] (\eta ) \, - \, D_x ( Q_\a (\eta ) ) \) \\
&=& \( [Q_\a,D_x] (\zeta_{(k)} ) \) \cdot (D_x \eta) \ - \
(D_x \zeta_{(k)} ) \cdot \( [Q_\a,D_x] (\eta ) \) \\
&=& \rho_\a^{\ \b} \ \[ \( Q_\b (\zeta_{(k)} )\)  \cdot (D_x \eta)
\ - \ (D_x \zeta_{(k)}) \cdot \( Q_\b (\eta) \) \]  \ .
\end{eqnarray*} We have used the hypothesis $Q_\a (\zeta_{(k)}) =
0 = Q_\a (\eta)$ for all $\a$. Using this once again, we conclude
that $\Theta_\a = 0$ for all $\a$; and hence that the IBDP holds
for $\chi$-prolongations, under the condition $\La = \la I$. \EOP
\bigskip

\section{The geometric meaning of twisted symmetries}
\label{sec:discussion}

We have seen that there are different types of twisted symmetries;
some of these enjoy the IBDP, and can then be used to perform
symmetry reduction of PDEs.

It is natural to wonder what is the characteristic singling out
the deformations of the prolongation operation which still allow
to perform reduction. The answer is actually simpler when we look
at combined twisted symmetries, as in the case of simple twisted
prolongations the situation is in some way a degenerate one and
some features are less apparent.

The condition for a vector field $X$ to be a symmetry (in its
different declinations) of a system $\De$ is that its prolongation
(standard or twisted) is tangent to the solution manifold $S_\De$;
in the same way, the condition for a system of vector fields
$\mathcal{X} = \{ X_\a \}$ to be a module (possibly an algebra) of
symmetries is that their prolongations (standard or twisted)
$\mathcal{Y} = \{ Y_\a \}$ are tangent to $S_\De$.

Note that the tangency condition is irrespective of a global
nonzero factor (possibly depending on the point in $J^n M$, i.e. a
smooth function $f : J^n M \to \R$). That is, if $Y$ is tangent to
$S_\De$, so is $f \cdot Y$; and similarly for systems of vector
fields.

We can reverse the point of view, and express the same tangency
condition putting the accent on $Y$ rather than on $S_\De$.

\medskip\noindent
{\bf Definition.} {\it A vector field $X$ (a system $\mathcal{X} =
\{ X_\a \}$ of vector fields) is a symmetry for $\De$ if $S_\De$
is an integral manifold for the distribution generated by the
prolongation $Y$ (the prolongations $\mathcal{Y} = \{ Y_\a \}$) of
$X$ (of the $X_\a$).}
\bigskip

It is obvious that this Definition is completely equivalent to the
standard one. However, focusing on distributions shows immediately
that we can change generators -- i.e. the vector field $Y$ or the
vector fields $Y_\a$ -- provided the distribution remains the
same.

In the case of a single generator $Y$, the only way to change
generator keeping the same distribution is by multiplying $Y$ by a
nowhere zero\footnote{We could actually allow for the factor to be
zero at the points where $Y$ is also zero; we disregard this
possibility for ease of writing, but the essential point (also
below) is that the rank of the distribution should remain the same
at each point.} factor $A (x,u,...,u_{(n)})$. If we want moreover
that the restriction of $Y$ to any one of the jet spaces $J^k M$
(for $k=0,...,n$) is well defined, i.e. that $Y$ is projectable in
$J^k M$ (in this case we say just it is {\it projectable} for
short), then we have to require that $A$ does not depend on the
derivatives of $u$, i.e. that $A = A(x,u)$.

If we look at vector fields $Y$ in $J^n M$ which are obtained in
this way from a vector field $Z$ which is a standardly prolonged
vector field (say the standard prolongation of $W$, a vector field
in $M$), i.e. $Y = A Z$, then $Y$ is the $\La$-prolongation of a
vector field in $M$. More precisely, we have the

\medskip\noindent
{\bf Lemma 14.} {\it Let $W$ be a vector field in $M$, and $Z$ its
standard prolongation to $J^n M$. Let $Y = \ga  Z$, with $\ga
(x,u)$ a nowhere zero function. Then $Y$ is the
$\La$-prolongation, with $\La = \la I$, of a vector field $X$ in
$M$, given by $X = \ga^{-1} W$; the functions $\ga$ and $\la$ are
related by \beq\label{eq:L14} \la \ = \ \ga \, (D_x \ga^{-1}) \ =
\ - \ (D_x \ga) \, \ga^{-1} \ . \eeq}

\medskip\noindent
{\bf Proof.} The proof actually mostly amounts to repeating (in
reverse) a computation already performed; see Lemma 4 and Remark 8,
but we give it for the sake of completeness.

Note that $Z$ is surely projectable, being the standard
prolongation of a Lie-point vector field; it follows immediately
that $Y$ is also projectable.  We write it in coordinates, with
the usual notation, as \beq Y \ = \ \^\xi \, \pa_x \ + \
\^\psi^a_{(k)} \, \pa_a^k \ . \eeq Here $\^\xi = \^\xi (x,u)$,
while $\^\psi^a_{(k)} = \^\psi^a_{(k)} (x,u,...,u_{(k)} )$, and we
also write $\^\psi^a_{(0)} = \^\phi^a$.

By assumption, $Z$ can be written as
$$ Z \ = \ \xi \, \pa_x \ + \ \psi^a_{(k)} \, \pa_a^k \ , $$
where the $\psi^a_{(k)}$ satisfy the prolongation formula
\eqref{eq:prolformODEs}, with $\psi^a_{(0)} = \phi^a$; this is the
prolongation of
$$ W \ = \ \xi \, \pa_x \ + \ \phi^a \, \pa_a \ . $$

We also know that $Y = \ga Z$, i.e.
$$ \^\xi \ = \ \ga \, \xi \ , \ \ \ \^\psi^a_{(k)} \ = \ \ga \,
\psi^a_{(k)} \ . $$ Therefore, using \eqref{eq:prolformODEs}, we
get
\begin{eqnarray*}
\^\psi^a_{(k+1)} &=& \ga \ \psi^a_{(k+1)} \\
&=& \ga \ \[ (D_x \psi^a_{(k)} ) \ - \ u^a_{(k+1)} \, (D_x \xi) \]
\\
&=& \ga \ \[ \( D_x (\ga^{-1} \^\psi^a_{(k)} ) \) \ - \
u^a_{(k+1)} \, \(D_x (\ga^{-1} \^\xi) \) \] \\
&=& \[ (D_x \^\psi^a_{(k)} ) \ - \ u^a_{(k+1)} \, (D_x \^\xi) \] \
+ \ \ga \, (D_x \ga^{-1}) \ \[  \^\psi^a_{(k)} \ - \ u^a_{(k+1)}
\, \^\xi \] \ . \end{eqnarray*} This means that the
$\^\psi^a_{(k)}$ obey the $\La$-prolongation formula
\eqref{eq:LAmbdaprol} with $$ \La \ = \ \[ \ga \, (D_x \ga^{-1})
\] \ I \ = \ - \ \[ (D_x \ga) \, \ga^{-1} \] \ I \ , $$ as stated
in the Lemma. \EOP

\medskip\noindent
{\bf Remark 27.} The fact that $\la$-prolongations are the {\it
only} vector fields\footnote{Of Lie-point type: a general classification would include contact symmetries; see \cite{PuS}.} which are collinear to (and hence admit the
same invariants as) a standardly prolonged vector field was first
noticed by Pucci and Saccomandi \cite{PuS} in a seminal paper which
set the basis for a geometrical understanding of $\la$-symmetries
and then of other types of twisted symmetries. \EOR
\bigskip

The following result is very simple but expresses the relations
between standard and $\la$-symmetries.

\medskip\noindent
{\bf Lemma 15.} {\it Let $X,Y,W,Z$ be as in Lemma 14 above. Then
$X$ is a $\La$-symmetry of the differential equations $\De$, with
$\La = \la I$, if and only if $W$ is a standard symmetry
of $\De$.}

\medskip\noindent
{\bf Proof.} The vector fields $Y$ and $Z$ span the same
distribution, hence the result follows from the Definition above.
\EOP
\bigskip

The results obtained above for simple twisted symmetries as
$\La$-symmetries  immediately generalizes (basically with the same
proof) to the case of collective twisted symmetries as
$\s$-symmetries.

\medskip\noindent
{\bf Lemma 16.} {\it Let $\mathcal{W} = \{ W_1 , ... , W_r \}$ be
a set of vector fields in involution in $M$, and $\mathcal{Z} = \{
Z_1 , ... , Z_r \}$ the set of their standard prolongations to
$J^n M$. Let $Y_\a = \Ga_\a^{\ \b} Z_\b$, with $\Ga = \Ga (x,u)$ a
nowhere singular matrix function on $M$. Then $\mathcal{Y}$ is the
$\s$-prolongation of a set of vector fields $\mathcal{X} = \{ X_1
, ... , X_r \}$ in $M$, given by $X_\a = (\Ga^{-1})_\a^{\ \b}
W_\b$; the matrix functions $G$ and $\s$ are related by
\beq\label{eq:L16} \s \ = \ \Ga \, (D_x \Ga^{-1} ) \ = \ - \, (D_x
\Ga) \, \Ga^{-1} \ . \eeq}

\medskip\noindent
{\bf Proof.} We proceed as in the proof to Lemma 14; again the
proof actually mostly amounts to repeating (in reverse) a
computation already performed, see Lemma 11.

The $Z_\a$ are surely projectable, and so are the $Y_\a$. We write
them, with the usual notation, as \beq Y_\a \ = \ \~\xi_\a \,
\pa_x \ + \ \~\psi^a_{\a,(k)} \, \pa_a^k \ . \eeq We also write
$\~\psi^a_{\a,(0)} = \~\phi^a_\a$.

By assumption, $Z_\a$ can be written as
$$ Z_\a \ = \ \xi_\a \, \pa_x \ + \ \psi^a_{\a,(k)} \, \pa_a^k \ , $$
with coefficients $\psi^a_{\a,(k)}$ satisfying the prolongation
formula \eqref{eq:prolformODEs}, with $\psi^a_{\a,(0)} =
\phi^a_\a$; this is the prolongation of
$$ W \ = \ \xi_\a \, \pa_x \ + \ \phi^a_\a \, \pa_a \ . $$

We also know that $Y = \Ga Z$, i.e.
$$ \~\xi_\a \ = \ \Ga_\a^{\ \b} \, \xi_\b \ , \ \ \ \~\psi^a_{\a,(k)} \ = \ \Ga_\a^{\ \b} \,
\psi^a_{\b,(k)} \ . $$ Therefore, using again
\eqref{eq:prolformODEs}, we get
\begin{eqnarray*}
\~\psi^a_{\a,(k+1)} &=& \Ga_\a^{\ \b} \ \psi^a_{\a,(k+1)} \\
&=& \Ga_\a^{\ \b} \ \[ (D_x \psi^a_{\b,(k)} ) \ - \ u^a_{(k+1)} \,
(D_x \xi_\b) \] \\
&=& \Ga_\a^{\ \eta} \ \[ \( D_x ((\Ga^{-1})_\eta^{\ \b}
\~\psi^a_{\b,(k)} ) \) \ - \
u^a_{(k+1)} \, \(D_x ((\Ga^{-1})_\eta^{\ \b} \~\xi_\b) \) \] \\
&=& \[ (D_x \~\psi^a_{\a,(k)} ) \ - \ u^a_{(k+1)} \, (D_x
\~\xi_\a)
\] \\ & & \ \ \ \ + \ \Ga_\a^{\ \eta} \, \(D_x (\Ga^{-1})_\eta^{\ \b}\)  \ \[  \~\psi^a_{\b,(k)} \ - \
u^a_{(k+1)} \, \~\xi_\b \] \ . \end{eqnarray*} This means that the
$\~\psi^a_{\a,(k)}$ obey the $\s$-prolongation formula
\eqref{eq:sigprol} with $\s$ as in \eqref{eq:L16}. \EOP

\medskip\noindent
{\bf Lemma 17.} {\it Let $\mathcal{X,Y,W,Z}$ be as in Lemma 16
above. Then $\mathcal{X}$ is a set of $\s$-symmetries of the
differential equations $\De$ if and only if $\mathcal{W}$
is a set of standard symmetries of $\De$.}

\medskip\noindent
{\bf Proof.} The set of vector fields $\{ Y_\a \}$ and $\{ Z_\a
\}$ span the same distribution, hence the result follows from the
Definition above. \EOP

\medskip\noindent
{\bf Remark 28.} One could also obtain similar results for
$\chi$-symmetries; however we have seen above that these are of
interest for symmetry reduction of ODEs only in the case where
$\La = \la I$; in this case $\chi$-prolongations with $\chi =
(\La,\^\s)$ are {\it de facto} equivalent to sigma-prolongations
with $\s = (\la I + \^\s)$, as seen also in the proof to Lemma 13.
\EOR
\bigskip

Finally, we will conclude our discussion with some words concerning the relation of these
results with the gauge-theoretic properties of twisted symmetries.

The fact that twisted symmetries are (locally) gauge-equivalent to
standard ones, or more precisely (and more generally) that vector
fields obtained via twisted prolongations are gauge equivalent to
vector fields obtained via standard prolongations, can now be
readily understood.

In fact, our Definition 1 focuses on distributions rather on
(their generating) vector fields; the gauge transformations which
appeared in discussing several types of twisted prolongations are
transformations which act on the generating sets, i.e. on vector
fields, changing them but keeping the distribution they generate
invariant. In this sense, the appearance of such gauge-theoretic
properties is entirely natural from the point of view of
distributions -- albeit it might seem rather mysterious when we
focus on single vector fields.

It should be stressed that our results concerning gauge equivalence of twisted and
standard symmetries were obtained under the assumption that the matrices
embodying the gauge transformations are invertible at all points; if this
assumption fails, the equivalence is only local, in open neighborhoods
not including singular points. The singular or nonsingular nature of the
concerned matrices is not left to our choice, but is commanded by the
function $\la$, or the matrix $\s$, or the matrices $\La_i$.

Actually, it is precisely the case where the gauge equivalence with standard
symmetries is only local to be the most interesting, in the sense in this case
the consideration of twisted symmetries allows to obtain really new results.\footnote{It
should also be recalled that, as discussed in Sect.\ref{sec:evolutionary},
the gauge equivalence can also connect twisted symmetries expressed by Lie-point
vector fields to standard symmetries (in the sense of symmetries arising from
standard prolongation) which are expressed by generalized, rather than Lie-point,
vector fields.}

\section{Other topics}
\label{sec:others}

We should finally mention some other relevant topics, also
investigated in relation to twisted symmetries, which have not
been discussed in the present paper, and for which we refer to the
literature.

First of all, twisted symmetries can also be formulated as a
generalization of {\it orbital symmetries}
\cite{HaWa,Wal99,WalSPT}; this is discussed in detail in
\cite{CGWjlt1}, see also \cite{Cic13,Sprol,SprolDS,CGWjlt2}.

In a very recent note \cite{Mor14}, Morando investigated the
relation with reducible structures, extending the approach
mentioned in Remark 3 above and exploring the relation with
Frobenius reduction.

The relation between twisted symmetries (in particular,
$\la$-symmetries), first integrals and integrating factors
\cite{BKS1,BKS2,Cic13,Cic13b,GO,Moh14,MuRom11,MuRom11b}, as well
as that between twisted symmetries (in particular,
$\la$-symmetries) and Jacobi Last Multiplier \cite{MuRom14,NL13}
has been studied by several authors .

Finally, we mention that twisted symmetries (in particular, once
again, $\la$-symmetries) and their applications have also been
studied for discrete equations \cite{LNR12,LR10}.

\section{Conclusions}

We have reviewed recent advances in twisted prolongations of
vector fields and twisted symmetries of (systems of) differential
equations, complementing our previous review paper \cite{Gtwist}
and providing an up-to-date discussion.

Recent advances concerned the introduction, beside simple ones, of
collective twisted prolongations. These in turn led to focus on
distributions rather than on single vector fields; this focus on
distributions -- even in the case of those generated by a single
vector fields, as is the case for  simple twisted symmetries -- helped in
turn in getting an understanding of the geometry of (collective,
but also simple) twisted prolongations, and explained why we have
local gauge equivalence between standard and twisted symmetries.



\end{document}